\documentclass[usenatbib]{mn2e}

\usepackage{url}
\usepackage{threeparttable}
\usepackage{graphicx}
\usepackage{longtable}

\usepackage{color}
\usepackage{ulem}
\usepackage{lscape}

\def\flux{erg~s$^{-1}$~cm$^{-2}$}
\def\ergs{erg~s$^{-1}$}

\title[Population of HMXBs in the Milky Way]{Population of persistent high mass X-ray binaries in the Milky Way}
\author[Lutovinov et al.]{A.A.~Lutovinov $^{1}$\thanks{E-mail:aal@iki.rssi.ru}, M.G.~Revnivtsev $^{1}$, S.S.~Tsygankov $^{2,3,1}$, R.A.~Krivonos  $^{4,1}$\\
$^{1}$ Space Research Institute, Russian Academy of Sciences, Profsoyuznaya 84/32, 117997 Moscow, Russia\\
$^{2}$ Finnish Centre for Astronomy with ESO (FINCA), University of Turku,  V\"ais\"al\"antie 20, FI-21500 Piikki\"o, Finland \\
$^{3}$ Astronomy Division, Department of Physics, FI-90014 University of Oulu, Finland\\
$^{4}$ Space Sciences Laboratory, 7 Gauss Way, University of California, Berkeley, CA 94720, USA
}
\begin{document}


\pagerange{\pageref{firstpage}--\pageref{lastpage}} \pubyear{2013}

\maketitle

\label{firstpage}

\begin{abstract}
We present results of the study of persistent high mass X-ray binaries (HMXBs) in the Milky Way, obtained from the deep INTEGRAL Galactic plane survey. This survey provides us a new insight into the population of high mass X-ray binaries because almost half of the whole sample consists of sources discovered with INTEGRAL.
It is demonstrated for the first time that the majority of persistent HMXBs have supergiant companions and their luminosity function steepens somewhere around $\sim2\times 10^{36}$ \ergs. We show that the spatial density distribution of HMXBs correlates well with the star formation rate distribution in the Galaxy. The vertical distribution of HMXBs has a scale-height $h\simeq85$ pc, that is somewhat larger than the distribution of young stars in the Galaxy. We propose a simple toy model, which adequately describes general properties of HMXBs in which neutron stars accrete a matter from the wind of the its companion (wind-fed NS-HMXBs population). Using the elaborated model we argue that a flaring activity of so-called supergiant fast X-ray transients, the recently recognized sub-sample of HMXBs, is likely related with the magnetic arrest of their accretion.
\end{abstract}


\section{Introduction}

A formation and evolution of stars and binary systems are very slow processes (typical timescales of $>10^6$ yr) and can not be directly traced. The only way to understand them is to study populations of sources. Properties of an ensemble of sources at different stages of their evolution do have characteristics which can be measured and compared with predictions of different models.

X-ray sources in our Galaxy were discovered quite long ago \cite[][]{giacconi62,giacconi71} and studied intensively in different ways during $\sim40$ years. Advances in a sensitivity and angular resolution of current X-ray telescopes allow now to detect and count such objects in a variety of outer galaxies \cite[see e.g.][]{trinchieri91,primini93,gilfanov04,kim04}.

In spite of this observational progress the properties of populations of different types of X-ray sources are not fully understood yet. The main reason for this is a scarce information about physical properties of large samples of HMXBs. Objects in outer galaxies can be counted and their X-ray appearances measured, but physical properties of detected sources (first of all, types and ages of their stellar components, orbital periods, etc.), which can help to understand their formation and evolution, rarely can be obtained (typically from optical or infrared instruments) due to their extreme faintness.

On the other hand -- physical properties of stellar binaries in our Galaxy can often be well measured, but the uniform highly sensitive surveys of such objects were absent until recently. A strong absorption of soft X-rays ($<2$ keV) in the interstellar medium or near binaries themselves have not allowed to obtain an unbiased sample of sources in the Galactic plane from the ROSAT all sky survey  \citep{voges99}. Surveys, performed with the ASCA \citep{sugizaki01} and XMM-Newton \citep{hands04} observatories, are somewhat free from these limitations, but both observatories covered only small areas of the Galactic plane.

X-ray instruments, which operate in the harder X-ray energy range, where the interstellar photo-absorption does not play an important role, and can cover the whole Galaxy, have typically a poor angular resolution and thus are affected by a source confusion in the Galactic plane (e.g., UHURU, \citealt{forman78}; RXTE, \citealt{markwardt00,revnivtsev04}). Samples of sources, collected over all history of X-ray astronomy \cite[e.g.,][]{liu07} can not to be taken as a representative for statistical and physical studies of populations due to their non-uniformity respect to the flux, detection criteria, etc.

A systematic survey of the Galaxy in 2003-2011 with the INTEGRAL observatory \citep{winkler03AA} in the hard X-ray energy range ($>17$ keV) with the moderate angular resolution ($\sim12^\prime$) allowed us for the first time to overcome all these difficulties and to perform virtually unbiased search for X-ray binaries in the Milky Way with an unprecedented sensitivity \citep{krivonos12}.

Previous observations of galactic sources with the INTEGRAL observatory have proved to be fruitful in a variety of fields: for explanation of the origin of the Galactic Ridge X-ray emission in hard X-rays \citep{krivonos07}; for systematic discoveries of strongly photoabsorbed high mass X-ray binaries and study of their distribution in the Galaxy \cite[see e.g.][]{courvoisier03,lutovinov05,2007A&A...467..585B,lutovinov07,chaty08,coleiro11,bodaghee12}; for confirmation of the presence of the break in the luminosity function of low mass X-ray binaries \citep{revnivtsev08a}, which is likely related with changes of the evolutionary type of stellar companions in these systems \citep{revnivtsev11}.

In this paper we present the most sensitive unbiased flux-limited sample of non-transient high mass X-ray binaries in our Galaxy. Our main goal is to establish and understand general properties of this population. This knowledge is a very important for different reasons: for understanding of properties of compact objects, their formation and evolution; for explanation of the observed behaviour of different types of HMXBs (in particular, supergiant fast X-ray transients); for interpretation of observations of variety of galaxies, which are now became possible with new generation of X-ray telescopes (in particular, to calculate a contribution of HMXBs to observed luminosities), etc.

\section{INTEGRAL survey of the Galactic plane. The sample of HMXBs}

In our study we use results from the INTEGRAL Galactic plane survey, presented in \citet{krivonos12}. The sensitivity of this survey is typically $10^{-11}$ \flux\ in the 17-60 keV energy band, which ensures the detection of sources with luminosities $\ga 10^{35}$ \ergs\ within a half of the Galaxy ($\la9$ kpc from the Sun) and $\ga5\times10^{35}$ \ergs\ over the whole Galaxy ($\la20$ kpc from the Sun).

One of advantages of the INTEGRAL Galactic plane survey is the multiple coverage of virtually all galactic sources. During the period of operation all parts of the Galaxy were observed tens of times, that allow us to quantify a long term behaviour of a majority of sources. This is important because the goal of our work is to understand properties of the population of high mass X-ray binaries, which are typically registered (or can be registered) in existing and future surveys both our Galaxy and distant galaxies. {\sl Therefore in our work we concentrate only on persistent sources}. Transiently appearing HMXBs typically have small duty cycles and thus they should be a minority among the population of HMXBs detected in snapshot observations of galaxies.

\begin{figure}
\includegraphics[width=\columnwidth,bb=49 149 550 700,clip]{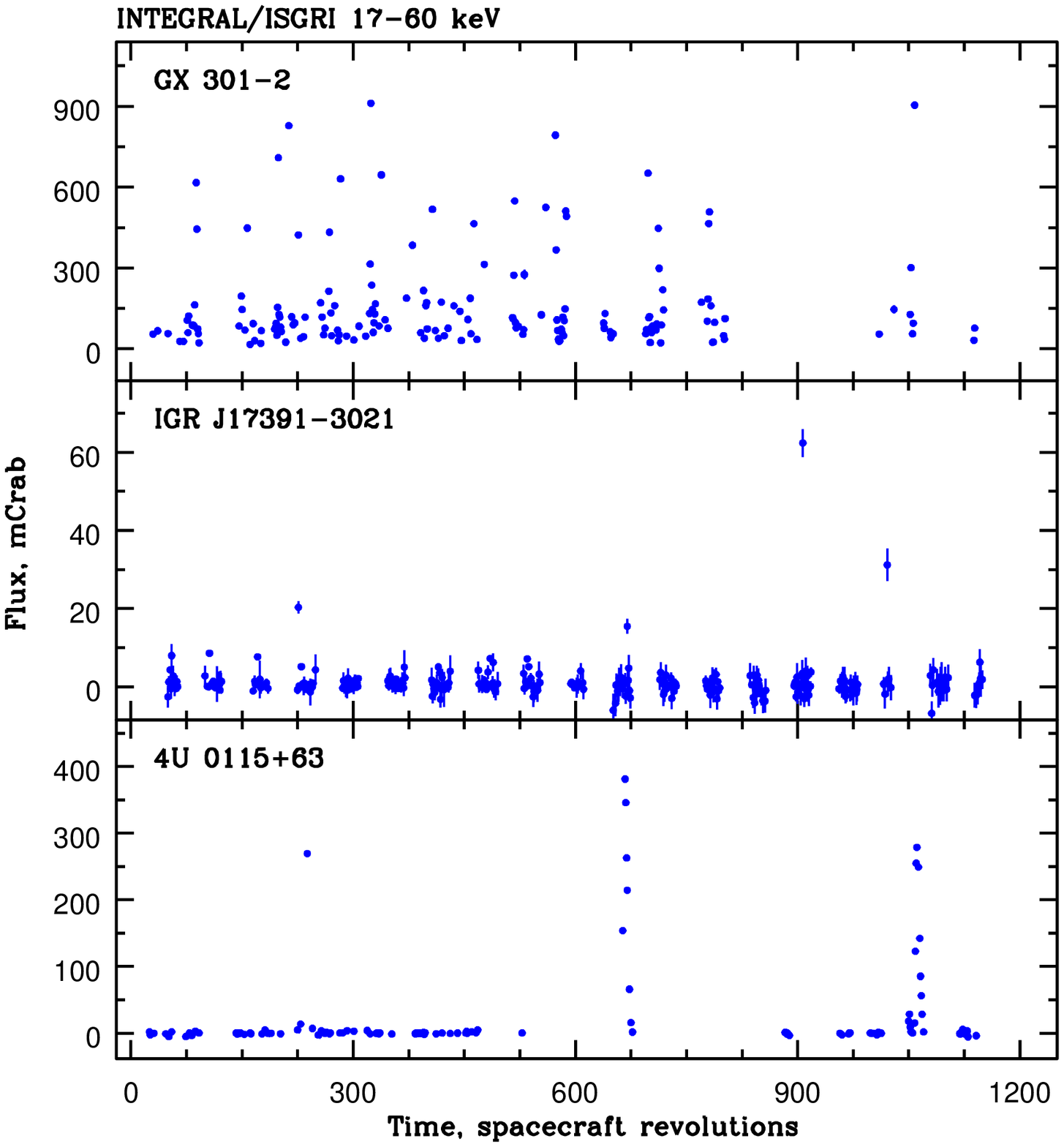}

\hspace{3mm}\includegraphics[width=0.96\columnwidth,bb=30 296 480 670,clip]{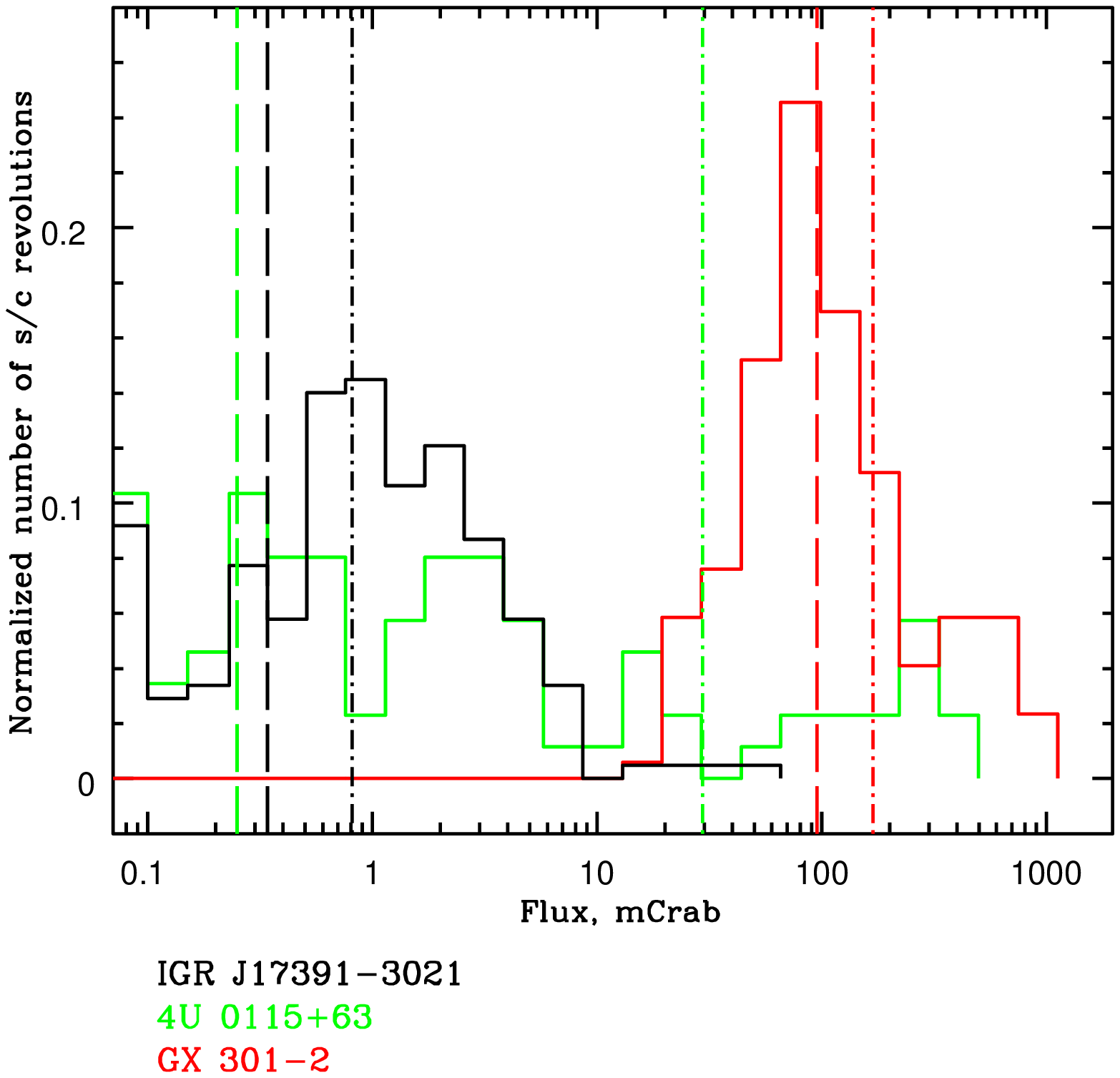}
\caption{{\it Three upper panels.} Examples of light curves of three sources of different types: persistent (GX\,301-2), supergiant fast X-ray transient (IGR J17391-3021) and Be-transient (4U\,0115+63). {\it Bottom panel.} Normalized distribution of values of fluxes, measured for these three sources in single revolutions (with the duration of about 3 days) of the INTEGRAL spacecraft (histograms). Dashed lines indicate median fluxes registered from sources, dashed-dotted lines -- mean flux values. Red color corresponds to the source GX\,301-2, black one -- to IGR\,J17391-3021, green one -- to 4U\,0115+63.}
\label{distr}
\end{figure}

\begin{figure}
\includegraphics[width=\columnwidth,bb=20 329 567 703,clip]{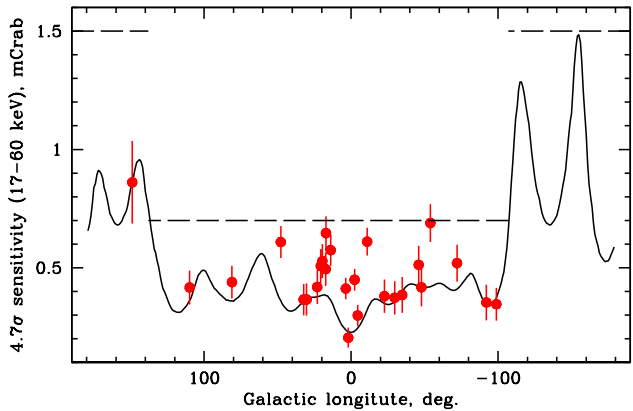}
\caption{Sensitivity of the INTEGRAL survey (at the $4.7\sigma$ level in the 17-60 keV energy band) over the whole Galaxy (solid line). Coordinates and fluxes of 26 non-identified persistent sources are shown by circles. Uncertainties correspond to $1\sigma$. Dashed line denotes the survey flux limit (see text for details).}
\label{sensflux}
\end{figure}

As a first step, among sources detected on the average map of the Galactic plane we exclude all sources which are known to be not high mass X-ray binaries. The remaining objects are either HMXBs or unidentified sources. In total, 75 HMXBs with a confirmed nature and 33 sources of an unknown origin (including HMXB candidates) at latitudes $|b|<5^\circ$ were detected during the INTEGRAL Galactic plane survey \citep{krivonos12}.

Then we exclude transient sources from our sample. There is no firmly established concept of a transiency, therefore this is a non-trivial task. To illustrate it, some examples of a different long term behaviour of sources are shown in Fig.\ref{distr}, where one can see light curves (upper panels) and distributions of fluxes (bottom panel) of three sources, representing three big families of high mass X-ray binaries: (quasi-)persistent sources, Be-transients and supergiant fast X-ray transients (SFXTs). We quantify the ''persistency'' of a source via a ratio of two flux values: the median flux value $F_{\rm median}$ and the mean flux value $\langle F\rangle$. For our main source sample we accept only sources which have the ratio $F_{\rm median}/\langle F \rangle >0.5$. This resulted in 54 confirmed HMXBs and 26 unidentified sources. It is necessary to note, that several SFXTs or candidates, which are named ''trantients'', still passed the selection criteria. It can be connected with a high duty cycle for some sources (like IGR\,J16479-4514) or due to a luminosity dynamic range not particularly high for some sources in the INTEGRAL energy band (this could be the case of the so called intermediate SFXTs IGR\,J17354-3255, IGR\,J16418-4532,  AX\,J1845.0-0433).

In order to further increase the identification completeness of our sample we raise the flux limit of the survey, as virtually all non-identified sources are faint. The sensitivity of the survey is shown as a function of the galactic longitude in Fig.\ref{sensflux}. It was calculated as the average value of the INTEGRAL/IBIS/ISGRI sensitivity over Galactic latitudes $-5^\circ<b<5^\circ$ for each longitude direction. Circles denote fluxes of non-identified sources with their uncertainties. Two important facts are clearly seen from this figure: 1) the survey sensitivity is not uniform over the Galaxy and is much better in its inner part (directions with $|l|\la100^\circ$); 2) all non-identified sources are faint. Thus, if we increase the survey flux limit up to $0.7$ mCrab ($10^{-11}$ \flux\ in the $17-60$ keV energy band) over the Galactic longitude range $-107^\circ<l<136^\circ$ and $1.5$ mCrab in the remaining part of the Galaxy there will be no non-identified sources in our sample. At the same time, the list of known HMXBs will shorten to 48 sources.

\begin{table*}

\caption{List of persistent galactic high-mass X-ray binaries detected by the INTEGRAL observatory\label{tab:srclist}}
\hspace{-7mm}\begin{tabular}{l|r|r|c|r|r|l|l}
\hline
\hline
\multicolumn{8}{c}{HMXBs with fluxes $>10^{-11}$ erg s$^{-1}$ cm$^{-2}$ and known distances} \\
\hline
Name & $l$, & $b$, & $L_{X, 17-60 keV}$, & Distance, & $P_{orb}$, & Class & References \\
& deg & deg & $10^{35}$ erg s$^{-1}$ & kpc & days & & \\
\hline
Vela X-1 & -96.93 & 3.93 & $5.868\pm0.003$ & 1.4 & 8.96 & B0.5Ib & 1, 2, 3 \\
3U 1022-55 & -74.64 & 1.49 & $0.311\pm0.033$ & 5.0 & & B0III-Ve & 4, 5 \\
Cen X-3 & -67.90 & 0.33 & $24.51\pm0.041$ & 5.7 & 2.09 & O6-7II-III & 6, 7, 8 \\
IGR J11305-6256 & -66.05 & -1.48 & $0.299\pm0.012$ & 3.0 & & B0IIIe & 9, 10 \\
IGR J11435-6109 & -65.12 & 0.68 & $3.165\pm0.097$ & 8.6 & 52.46 &B2III or B0V& 11, 12 \\
A 1145.1-6141 & -64.50 & -0.02 & $20.14\pm0.095$ & 8.5 & 14.4 & B2Iae & 13, 14 \\
X 1145-619 & -64.38 & -0.24 & $0.271\pm0.012$ & 3.1 & 187.5 & B1Vne & 15, 5, 16 \\
1ES 1210-646 & -61.13 & -2.31 & $0.105\pm0.011$ & 2.8 & & B2V & 11 \\
GX 301-2 & -59.90 & -0.03 & $31.53\pm0.016$ & 3.5 & 41.5 & B1Ia+ & 17, 18, 19 \\
1RXP J130159.6-635806 & -55.91 & -1.12 & $0.765\pm0.041$ & 5.5 & &O9V or B1III& 20, 9 \\
4U 1416-62 & -46.98 & -1.57 & $0.438\pm0.048$ & 6.0 & 42.12 & B1Ve & 21, 5, 22 \\
4U 1538-522 & -32.58 & 2.16 & $5.093\pm0.024$ & 4.5 & 3.73 & B0.2Ia & 23, 24, 25 \\
IGR J16207-5129 & -27.54 & -1.05 & $1.639\pm0.041$ & 6.1 & 9.73 & O7.5 & 26, 27, 28 \\
& & & & & & B1Ia & 26 \\
IGR J16195-4945 & -26.44 & 0.33 & $0.494\pm0.022$ & 4.5 & & B1sg & 29, 27 \\
IGR J16318-4848 & -24.38 & -0.44 & $0.890\pm0.002$ & 1.6 & & sgB[e] & 27 \\
IGR J16320-4751 & -23.67 & 0.16 & $2.995\pm0.013$ & 3.5 & 8.96 & O8I & 27, 30 \\
AX J163904-4642 & -21.99 & 0.07 & $7.964\pm0.125$ & 10.6 & & BIV-V & 31 \\
IGR J16418-4532 & -20.81 & 0.49 & $9.677\pm0.191$ & 13.0 & 3.75 & O8.5(sg?) & 31, 27, 32 \\
IGR J16465-4507 & -19.94 & 0.13 & $1.714\pm0.100$ & 9.4 & 30.32 & B0.5I & 27, 33 \\
& & & & & & O9.5Ia & 26 \\
IGR J16479-4514 & -19.84 & -0.12 & $0.445\pm0.009$ & 2.8 & 3.32 & O8.5I & 26, 27, 34 \\
& & & & & & O9.5Iab & 26 \\
IGR J16493-4348 & -18.62 & 0.57 & $6.819\pm0.260$ & 15.0 & 6.78 & B0.5Ib & 35, 36, 37 \\
OAO 1657-415 & -15.63 & 0.32 & $48.72\pm0.058$ & 7.1 & 10.4 & B0-6sg & 38, 39, 40 \\
4U 1700-377 & -12.24 & 2.17 & $12.98\pm0.004$ & 2.12 & 3.41 & O6.5Iaf+ & 41, 42, 43 \\
EXO 1722-363 & -8.50 & -0.35 & $3.996\pm0.032$ & 6.1 & 9.74 & B0-B1 Ia & 27, 44, 45 \\
AX J1749.1-2733 & 1.58 & 0.06 & $3.285\pm0.128$ & 13.5 & & B1-3 & 46 \\
AX J1749.2-2725 & 1.70 & 0.11 & $2.760\pm0.137$ & 14.0 & & B3 & 46 \\
IGR J18027-2016 & 9.43 & 1.03 & $10.24\pm0.140$ & 12.4 & 4.6 & B1b & 47, 48 \\
IGR J18214-1318 & 17.67 & 0.48 & $1.377\pm0.076$ & 8.0 & & B0V-O9I & 49 \\
AX J1845.0-0433 & 28.14 & -0.66 & $0.255\pm0.015$ & 3.6 & & O9.5I & 50 \\
XTE J1855-026 & 31.07 & -2.09 & $14.44\pm0.114$ & 10.0 & 6.07 & B0Iaep & 51, 52, 53 \\
X 1908+075 & 41.89 & -0.81 & $9.658\pm0.047$ & 7.0 & 4.4 & O7.5-9.5sg & 54, 55 \\
4U 1907+097 & 43.74 & 0.47 & $4.389\pm0.024$ & 5.0 & 8.38 & O8-9Ia & 56, 56, 57 \\
IGR J19140+0951 & 44.29 & -0.46 & $1.687\pm0.012$ & 3.6 & 13.56 & B1I & 47, 27, 58 \\
& & & & & & B0.5I & 26 \\
SWIFT J2000.6+3210 & 68.98 & 1.13 & $2.108\pm0.092$ & 8.0 & &early BV or mid BIII & 59 \\
4U 2206+543 & 100.60 & -1.10 & $0.852\pm0.010$ & 2.6 & 9.57 & O9.5V & 60, 61 \\
1A 0114+650 & 125.71 & 2.55 & $6.642\pm0.063$ & 7.2 & 11.6 & B1Ia & 62, 63 \\
RX J0146.9+6121 & 129.52 & -0.80 & $0.096\pm0.010$ & 2.5 & & B1Ve & 64, 5 \\[1mm]
\hline
\multicolumn{8}{c}{HMXBs and candidates to HMXBs with fluxes $>10^{-11}$ erg s$^{-1}$ cm$^{-2}$ and unknown distances} \\
\hline
IGR J10100-5655 & -77.76 & -0.67 & $0.713\pm0.086^{a}$ & & & early giant& 70 \\
AX J1700.2-4220 & -16.23 & -0.03 & $1.124\pm0.070^{a}$ & & 44.03 & Be & 71 \\
IGR J17200-3116 & -4.99 & 3.34 & $1.500\pm0.048^{a}$ & & & HMXB & 70 \\
IGR J17354-3255 & -4.54 & -0.26 & $0.958\pm0.045^{a}$ & & 8.448 & HMXB & 72, 73 \\
IGR J17586-2129 & 8.01 & 1.36 & $1.502\pm0.052^{a}$ & & & HMXB & 74 \\[1mm]
\hline
\multicolumn{8}{c}{HMXBs with fluxes $<10^{-11}$ erg s$^{-1}$ cm$^{-2}$} \\
\hline
IGR J00370+6122 & 121.21 & -1.42 & $0.084\pm0.010$ & 3.0 & 15.665 &BN0.5II-IIIb& 68, 69 \\
IGR J14331-6112 & -45.15 & -0.76 & $1.053\pm0.135$ & 10.0 & & BIII or BV & 59 \\
IGR J16283-4838 & -24.66 & 0.08 & $3.032\pm0.339$ & 17.6 & & OBsg & 65 \\
IGR J17544-2619 & 3.24 & -0.32 & $0.130\pm0.010$ & 3.6 & 4.93 & O9Ib & 27, 66, 67 \\
IGR J18410-0535 & 26.78 & -0.23 & $0.113\pm0.011$ & 3.2 & & B1Ib & 26 \\
IGR J22534+6243 & 109.87 & 2.88 & $0.417\pm0.072^{a}$ & & & HMXB & 75 \\[1mm]
\hline
\end{tabular}
\end{table*}
\begin{table*}
{\bf Table 1.} (continue)

\begin{tabular}{l|r|r|c|r|r|l|l}
\hline
\hline
\multicolumn{8}{c}{HMXBs with black holes} \\
\hline
Name~~~~~~~~~~~~~~~~~~~~~~~~~~ & $l$, & $b$, & $L_{X, 17-60 keV}$, & Distance, & $P_{orb}$, & Class~~~~~~~~~~~~~~~~~~ & References \\
& deg & deg & $10^{35}$ erg s$^{-1}$ & kpc & days & & \\
\hline
SS 433 & 39.69 & -2.24 & $3.786\pm0.030$ & 5.5 & 13.1 & Asg & 76, 77 \\
Cyg X-1 & 71.34 & 3.07 & $38.96\pm0.005$ & 1.86 & 5.6 & O9.7Iab & 78, 79, 80 \\
Cyg X-3 & 79.85 & 0.70 & $101.2\pm0.057$ & 7.2 & 0.2 & WR & 81, 82, 83 \\[1mm]
\hline
\multicolumn{8}{c}{$\gamma$-loud HMXBs} \\
\hline
PSR B1259-63 & -55.82 & -0.99 & $0.119\pm0.010$ & 2.3 & 1236.7 & B2e & 84, 85, 86 \\
LS 5039 & 16.87 & -1.29 & $0.075\pm0.007$ & 2.9 & 3.906 & O6.5Vf & 87, 88, 89 \\
LSI 61 +303 & 135.67 & 1.07 & $0.143\pm0.013$ & 2.0 & 26.496 & B0Ve & 90, 91, 92 \\
\hline
\end{tabular}
\medskip
\begin{tabular}{ll}
$^{a}$ & flux in the 17-60 keV energy band, in mCrabs \\
\end{tabular}
\medskip

\parbox{17.5cm}
{\tiny
References:
(1) \cite{1998A&A...330..201C},
(2) \cite{1972ApJ...175L..19H},
(3) \cite{1984ApJ...283L..53B},
(4) \cite{1997A&A...323..853M},
(5) \cite{2011Ap&SS.332....1R},
(6) \cite{2009ApJ...691.1744T},
(7) \cite{1999MNRAS.307..357A},
(8) \cite{1972ApJ...172L..79S},
(9) \cite{2006A&A...449.1139M},
(10) \cite{2008ApJ...685.1143T},
(11) \cite{2009A&A...495..121M},
(12) \cite{2005ATel..377....1C},
(13) \cite{2002ApJ...581.1293R},
(14) \cite{1982MNRAS.201..171D},
(15) \cite{1997MNRAS.288..988S},
(16) \cite{1981AJ.....86..871H},
(17) \cite{2006A&A...457..595K},
(18) \cite{1997ApJ...479..933K},
(19) \cite{1986ApJ...304..241S},
(20) \cite{2007A&A...467..585B},
(21) \cite{1984ApJ...276..621G},
(22) \cite{1996A&AS..120C.209F},
(23) \cite{2004ApJ...610..956C},
(24) \cite{1978MNRAS.184P..73P},
(25) \cite{2000ApJ...542L.131C},
(26) \cite{2008A&A...486..911N},
(27) \cite{2008A&A...484..801R},
(28) \cite{2011ATel.3785....1J},
(29) \cite{2005A&A...429L..47S},
(30) \cite{2005ATel..649....1C},
(31) \cite{2008A&A...484..783C},
(32) \cite{2006ATel..779....1C},
(33) \cite{2010MNRAS.406L..75C},
(34) \cite{2009MNRAS.397L..11J},
(35) \cite{2010A&A...516A.106N},
(36) \cite{2008ATel.1396....1N},
(37) \cite{2010MNRAS.406L..16C},
(38) \cite{2006MNRAS.367.1147A},
(39) \cite{2002ApJ...573..789C},
(40) \cite{1993ApJ...403L..33C},
(41) \cite{2009A&A...507..833M},
(42) \cite{2002A&A...392..909C},
(43) \cite{1973ApJ...184L..65J},
(44) \cite{2009A&A...505..281M},
(45) \cite{2003ATel..179....1M},
(46) \cite{2010MNRAS.409L..69K},
(47) \cite{2010A&A...510A..61T},
(48) \cite{2003ApJ...596L..63A},
(49) \cite{2009ApJ...698..502B},
(50) \cite{1996MNRAS.281..333C},
(51) \cite{1999ApJ...517..956C},
(52) \cite{2008ATel.1876....1N},
(53) \cite{2002ApJ...577..923C},
(54) \cite{2005MNRAS.356..665M},
(55) \cite{2000ApJ...532.1119W},
(56) \cite{2005A&A...436..661C},
(57) \cite{1980MNRAS.193P...7M},
(58) \cite{2004ATel..269....1C},
(59) \cite{2008A&A...482..113M},
(60) \cite{2006A&A...446.1095B},
(61) \cite{2001ApJ...562..936C},
(62) \cite{1996A&A...311..879R},
(63) \cite{1985ApJ...299..839C},
(64) \cite{1991MNRAS.253..649T},
(65) \cite{2011A&A...526A..15P},
(66) \cite{2006A&A...455..653P},
(67) \cite{2009MNRAS.399L.113C},
(68) \cite{2004ATel..285....1N},
(69) \cite{2004ATel..281....1D},
(70) \cite{2006A&A...459...21M},
(71) \cite{2010ATel.2564....1M},
(72) \cite{2009ATel.2022....1T},
(73) \cite{2011A&A...529A..30D},
(74) \cite{2009ApJ...701..811T},
(75) \cite{2012ATel.4248....1M},
(76) \cite{2002ApJ...578L..67G},
(77) \cite{1984ARA&A..22..507M},
(78) \cite{2011ApJ...742...83R},
(79) \cite{1973ApJ...179L.123W},
(80) \cite{1999MNRAS.302L...1P},
(81) \cite{2009ApJ...695.1111L},
(82) \cite{1992Natur.355..703V},
(83) \cite{1973NPhS..241...28C},
(84) \cite{2011ApJ...732L..11N},
(85) \cite{1994MNRAS.268..430J},
(86) \cite{2004MNRAS.351..599W},
(87) \cite{2012A&A...543A..26M},
(88) \cite{2001A&A...376..476C},
(89) \cite{2005MNRAS.364..899C},
(90) \cite{1991AJ....101.2126F},
(91) \cite{2005MNRAS.360.1105C},
(92) \cite{2002ApJ...575..427G}.
}
\end{table*}


This sample contains 37 persistent sources with accreting neutron stars and known distances, five HMXBs with unknown distances and six sources of a peculiar nature (HMXBs harboring black holes and $\gamma$-loud HMXBs).

The list of selected persistent HMXBs (and non-identified) sources with their main parameters is presented in Table \ref{tab:srclist}; its contents are described below.

{\it Column (1) ``Name''} -- source name.

{\it Columns (2,3) ``$l$, $b$''} -- source galactic coordinates, longitude and latitude, respectively.

{\it Column (4) ``Luminosity''} -- time-averaged source luminosity in the $17-60$ keV energy band. For sources with unknown distances the flux in mCrabs in this energy band is mentioned.

{\it Column (5) ''Distance''} -- distance to the source in kpc.

{\it Column (6) ''$P_{orb}$''} -- orbital period of the system.

{\it Column (7) ''Class''} -- optical class of the normal companion in the binary system. In some cases two possible type are indicated.

{\it Column (8) ``References, notes''} -- references and alternative source names. References correspond to the distance, class and orbital period measurements.\\

\begin{figure*}

\includegraphics[width=23cm,bb=20 20 575 330,angle=90]{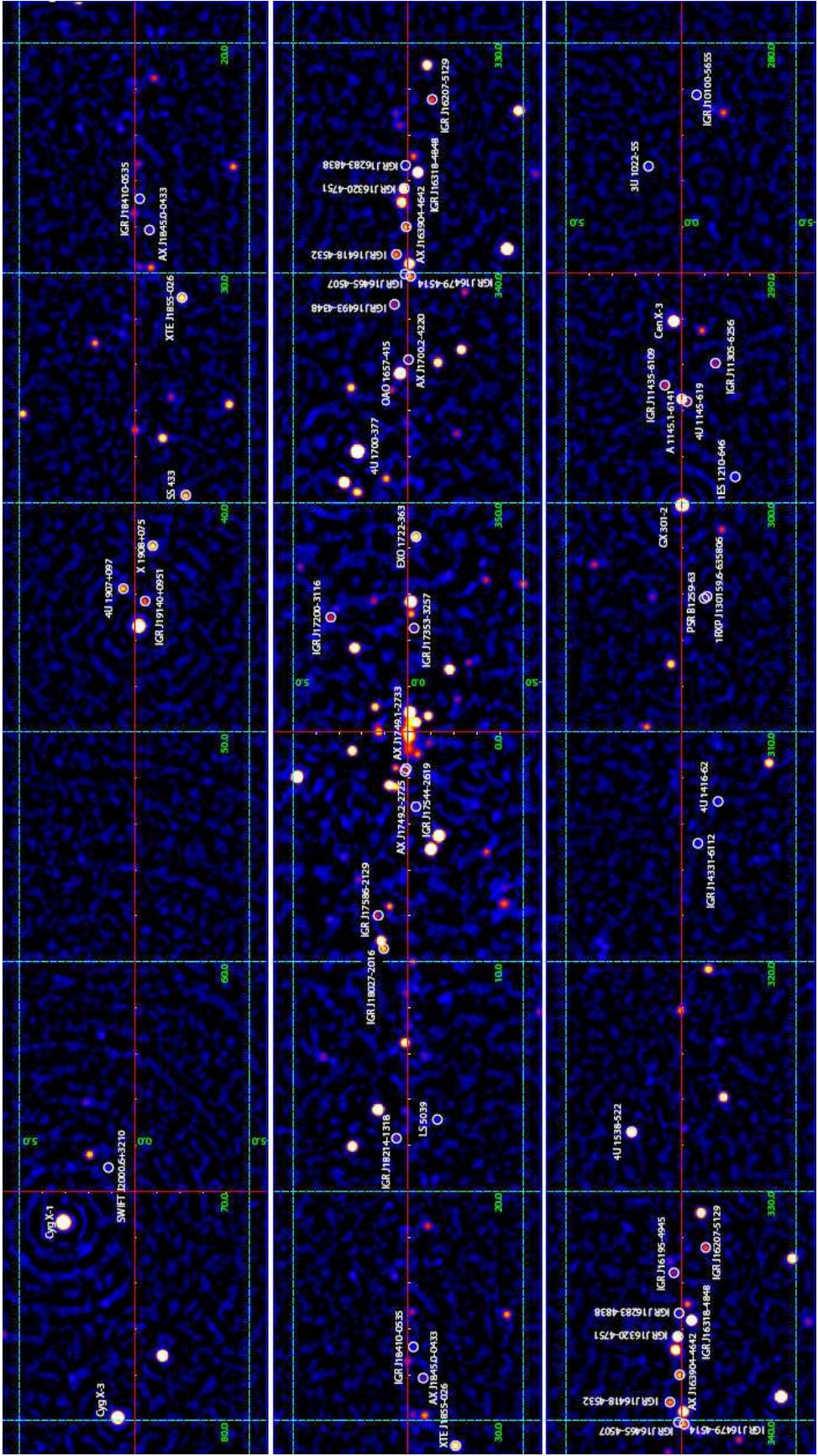}

\caption{Maps of the inner part of the Galactic plane, obtained with INTEGRAL/IBIS in the 17-60 keV energy band. All persistent sources from Table \ref{tab:srclist} are marked with circles and names.} \label{maps}
\end{figure*}

The spatial distribution of HMXBs from Table \ref{tab:srclist} are shown in Fig.\ref{maps}.

\section{Luminosity function and spatial distribution of wind-fed HMXBs}

In our Galaxy there are (and it can be seen from the Table\ref{tab:srclist}) several types of persistent X-ray binaries with massive companions. The most numerous population of them is binaries, in which the neutron star accrete a matter from the stellar wind of the massive companion (wind-fed systems). Other types of persistent X-ray binaries with massive stars include:
\begin{itemize}
\item binaries in which neutron stars accrete matter due to the Roche lobe overflow of the giant companion star (likely Cen X-3, see e.g. \citealt{lamers76});
\item binaries with accreting black holes (Cyg\,X-1, likely Cyg\,X-3);
\item binaries with a super-Eddington regime of accretion, in which the central engine is obscured by the accretion disk and only jet is visible in X-rays (e.g., SS\,433,  \citealt{fabrika04};
\item binaries where an X-ray (and gamma-ray) emission originates as a result of a non-thermal emission of particles accelerated in colliding winds of components (PSR\,B1259-63, LS\,5039, LSI\,+61\,303, Eta Carinae, etc.).
\end{itemize}

These classes of sources have only a few representatives in our Galaxy, which makes a study of their statistics impossible. {\sl Therefore we will concentrate in this paper only on binary systems with the wind-fed accreting neutron stars.}

A distribution of HMXBs over their luminosities (luminosity function, LF) is the simplest, but still an informative characteristic, which can be calculated from the sample of sources. It can be easily done for outer galaxies, but in the case of the Milky Way, while calculating the HMXBs LF, we should make an appropriate correction for an incomplete coverage of the Galaxy at different X-ray luminosities. In other words -- we should correct for the fact that sources with given luminosities are detectable for the INTEGRAL survey only within some distance limits.

The simplest way to do such a correction was adopted by \citet{grimm02} and \citet{voss10}, where authors assumed some particular volume density distribution of HMXBs over the Galaxy. In the former work it was assumed that the HMXBs volume density distribution has a disk shape with certain parameters, in the latter paper authors suggested that HMXBs are distributed similar to the stellar mass in the Galaxy.

In this work we want to measure the HMXBs density distribution over the Galaxy rather then assume it. Therefore we have adopted here a different approach.

\subsection{Volume limited samples}

\begin{figure}
\includegraphics[width=\columnwidth,bb=28 180 580 695,clip]{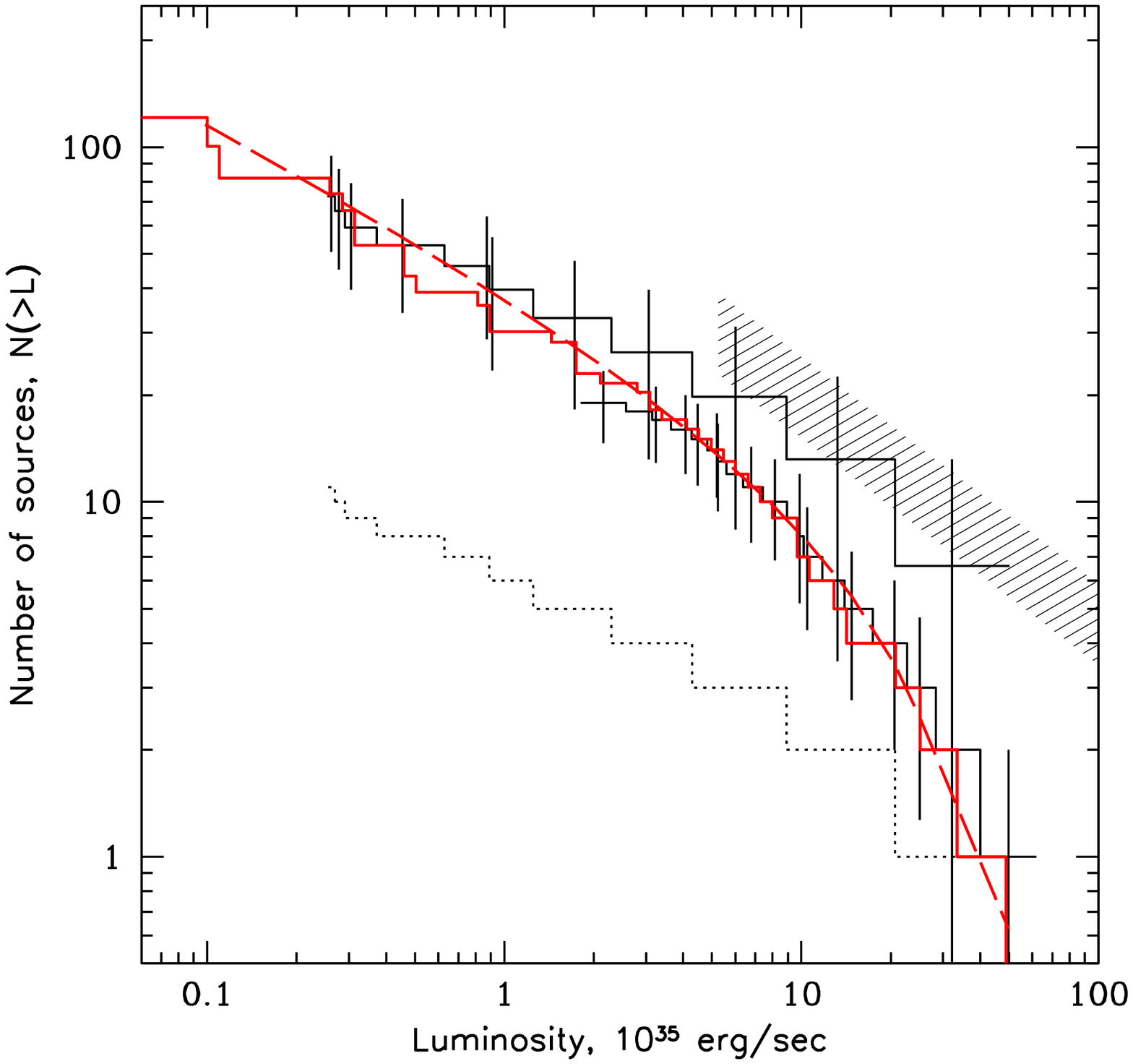}
\caption{Luminosity functions of the wind fed accreting HMXBs in the Galaxy. Two black solid histograms represent luminosity functions within volume limited samples (see text for details). Dotted histogram is the original number-luminosity function of the volume limited sample of sources within $d<4.1$ kpc. This function was multiplied by a correction factor $6.6$, calculated from densities of HMXBs in different galactocentric annuli. Red solid histogram is a luminosity function of the whole sample (normalized to the number of HMXBs over the whole Galaxy), calculated taking into account luminosity dependent corrections (due to limited sensitivity of the survey), dashed red curve -- the best fit model of the luminosity function with parameters from Table 2. Hatched area shows the number-luminosity function of all classes of HMXBs in our Galaxy from \citet{grimm02}.}
\label{lognlogs}
\end{figure}

As a first step we divided the whole available luminosity range of HMXBs into two intervals -- above the luminosity $2\times10^{35}$ \ergs\ (taking into account the adopted flux limit of our survey $10^{-11}$ \flux\ this sample is complete up to $\simeq13$ kpc from the Sun) and above the luminosity $2\times10^{34}$ \ergs\ (complete up to $\simeq4.1$ kpc from the Sun). For such volume-limited samples it is not needed to make a luminosity dependent correction and the luminosity distribution can be calculated directly. Results of these calculations -- two portions of the luminosity function of HMXBs -- are shown in Fig.\ref{lognlogs} by histograms. For the faint part of the luminosity function we present both observed number of sources in the volume limited sample (at distances $<4.1$ kpc from the Sun, dotted line) and the same histogram multiplied by a factor of $6.6$, which is a re-normalization, calculated from the measured density distribution of HMXBs (see below).

This plot provides us an indication that the luminosity function of HMXBs in the whole luminosity range ($10^{34}-10^{37}$ \ergs) is not following a single power law, but rather is curved at luminosities around $(0.4-2)\times10^{36}$ \ergs. The maximum likelihood approximation of the LF with simple power law functions at $L>4\times10^{35}$ \ergs\ and $L>2\times10^{34}$ \ergs\ gives the best fit values of their slopes $\gamma_{\rm faint}=1.49\pm0.21$ ($34.3<\log L_x<36.5$) and $\gamma_{\rm bright}=2.0\pm0.3$ ($35.6<\log L_x<36.5$). From a purely statistical point of view this difference in slopes is $\sim 2\sigma$ significant.

It is important to emphasize that distances to sources (Table \ref{tab:srclist}) are known with a limited accuracy due to different reasons. In general, it is reasonable to assume that they have an average accuracy not worse than $\sim$20\%. It means that the luminosity of HMXBs will be uncertain with a factor of $\sim$40\% for the known average flux values. To estimate a possible influence of these uncertainties on our results we have performed simple simulations. We have varied distances around values, presented in Table \ref{tab:srclist}, assuming a Gaussian distribution with $\sigma$=20\% of the source distance. Systematic uncertainties for the LF slopes calculated by this way are: $\Delta \gamma_{\rm faint}\approx0.06$, $\Delta \gamma_{\rm bright}\approx0.1$, respectively, i.e. well within the statistical uncertainties.

\subsection{Luminosity function of the whole sample}

In order to calculate the luminosity function of the whole sample of HMXBs we should take into account the fact that sources with given luminosities are detectable for the INTEGRAL survey only within some distance limits. In order to do this we need to determine their surface density distribution.

Given the limited accuracy of distances to sources in our sample we have limited ourself by an axially symmetric distribution of HMXBs in the Galaxy. We have divided the Galaxy into annuli with radii $R_{\rm g}<2$ kpc, $2-5$ kpc, $5-8$ kpc, $8-11$ kpc, $11-14$ kpc from the center (the distance from the Sun to the Galactic center is expected to be 8.5 kpc).

We assume that:

\begin{itemize}
\item the surface density of HMXBs (src/kpc$^{-2}$) is constant within each annulus;
\item shape of the luminosity function and its parameters are the same for all annuli.
\end{itemize}

In our flux limited survey the sources with different luminosities can be detected within different galactic areas. In order to take this effect into account we have adopted method $1/V_{\rm max}$ \citep{schmidt68}. For estimation of parameters of the HMXBs luminosity function $\phi(L)=dN/dL$ we used the Cash-statistics \citep{cash79} as follows:

\begin{eqnarray}
C=2\sum_j \left( \int\phi(L) S_{{\rm max},j}(L)dL - \sum_{i=1}^{N_j}\ln{ \left[\phi(L_{i,j})S_{{\rm max},j}(L_{i,j})\right]} \right) \nonumber
\end{eqnarray}
\hfill (1)
\medskip

Here, a summation $j$ goes over the set of annuli and a summation $i$ goes over $N_j$ sources within each annulus. $S_{{\rm max},j}$ is the maximum area of annulus $j$, within which a source with the luminosity $L_{i,j}$ can be detected.

A minimization of the $C$-statistics gives us best fit parameters of the luminosity function and its normalization in each annulus. We have adopted a simple broken power law shape of the luminosity function with slopes $\alpha_1$ and $\alpha_2$ below and above the break at the luminosity $L_{*}$:

$$
{dN\over{dL}}= \left\{ \begin{array}{rl}
A_{j}(L/{L_*})^{-\alpha_1}&\mbox{ if $L<L_*$}\\
A_{j}(L/{L_*})^{-\alpha_2}&\mbox{ if $L>L_*$}  ~~~~~~~~~~~~~~~~~~~~~~~~~~ (2)
\end{array} \right.
$$
where $A_j$ -- normalization of the luminosity function in each annulus $j$. Best fit parameters of this model with uncertainties are presented in Table 2. Statistical uncertainties correspond to a $1\sigma$ level; the systematical ones were calculated by variations of distances to sources as it was described in Section 3.1.

The shape of the LF of wind-fed HMXBs demonstrates a break, or at least a curvature. From a purely statistical point of view the statistical significance of the break is $\sim5$\%, or $\sim2\sigma$ ($\Delta C\approx5.9$ for 2 additional parameters). Obviously, this can not be interpreted as a solid detection, but rather as a possible indication of its presence. It should be noted additionally, that evidences for some flattening of the HMXBs LF was mentioned earlier by \cite{voss10} for sources in our Galaxy and \cite{shtykovsky05} for the population of HMXBs in the Small Magellanic Cloud. Thus it is likely that the gradual flattening of the HMXBs LF is real.

\begin{table}
\label{lfpars}
\caption{Best fit parameters of the luminosity function of HMXBs and their spatial density distribution}
\begin{tabular}{cc}
\hline
\hline
Parameter&Value and $1\sigma$ error\\
\hline
$\alpha_1$&$1.40\pm0.13$(stat.)$\pm0.06$(syst.)\\[1mm]
$\alpha_2$&$>2.2$\\[1mm]
$L_*, 10^{36}$ \ergs&$2.5^{+2.7}_{-1.3}$(stat.)$\pm1.0$(syst.)\\[1mm]
\hline
$R_{\rm g}$, kpc&$N(L>10^{35}$ \ergs) kpc$^{-2}$\\
\hline
0-2      & $0.0\pm0.05$(syst.)\\[1mm]
2-5      & $0.11^{+0.05}_{-0.04}$(stat.)$\pm0.02$(syst.)\\[1mm]
5-8      & $0.13^{+0.04}_{-0.03}$(stat.)$\pm0.01$(syst.)\\[1mm]
8-11    & $(3.8^{+2.1}_{-1.2})\times10^{-2}$(stat.)$\pm6.5\times10^{-3}$(syst.)\\[1mm]
11-14  & $(6.2^{+7.2}_{-4.3})\times10^{-3}$(stat.)$\pm4.8\times10^{-3}$(syst.)\\[1mm]
\hline
\end{tabular}
\end{table}

\subsection{Radial distribution}

The HMXBs surface densities (averaged over corresponding annuli) are presented in Table 2 and Fig. 5,6. It can be seen that the overall distribution of surface density of HMXBs in the Galaxy has a peak at galactocentric radii $2-8$ kpc.

Typical ages of high mass X-ray binaries are not longer then tens of Myrs, therefore it is natural to anticipate that their spatial density traces regions of the recent star formation rather then the stellar mass distribution \cite[see, e.g.,][]{haberl00}. Previous measurements of the HMXBs distribution in our Galaxy \cite[see e.g.][]{grimm02,lutovinov05,lutovinov07,bodaghee12} supports this general conclusion, but various incompleteness of previously available samples of HMXBs precluded accurate estimates of their global density distribution.

A comparison of the obtained HMXBs surface densities with the star formation surface densities taken from papers of \citet{guesten82,lyne85,chiappini01} shows their very good correlation: $N(HMXB, L_{\rm x}>10^{35}\textrm{\ergs})$/kpc$^2\approx 5.5\times 10^{-2} ~SFR/SFR_{\odot}$ (see Fig.\ref{sfr_radius}), here $SFR_{\odot}$ is the surface density of the star formation rate near the Sun.

It is necessary to note, that on the average map, obtained with the INTEGRAL observatory, there are 5 HMXBs with fluxes $>0.7$ mCrab, but with unknown distances (see Table \ref{tab:srclist}). These sources were not included into the calculation of the HMXBs surface densities and the luminosity function. This means that the true surface densities of HMXBs in the inner rings of the Galaxy (where most of sources are concentrated) might be slightly ($\sim14\%$) higher than presented in Table 2.

\begin{figure}
\includegraphics[width=\columnwidth,bb=43 170 530 630,clip]{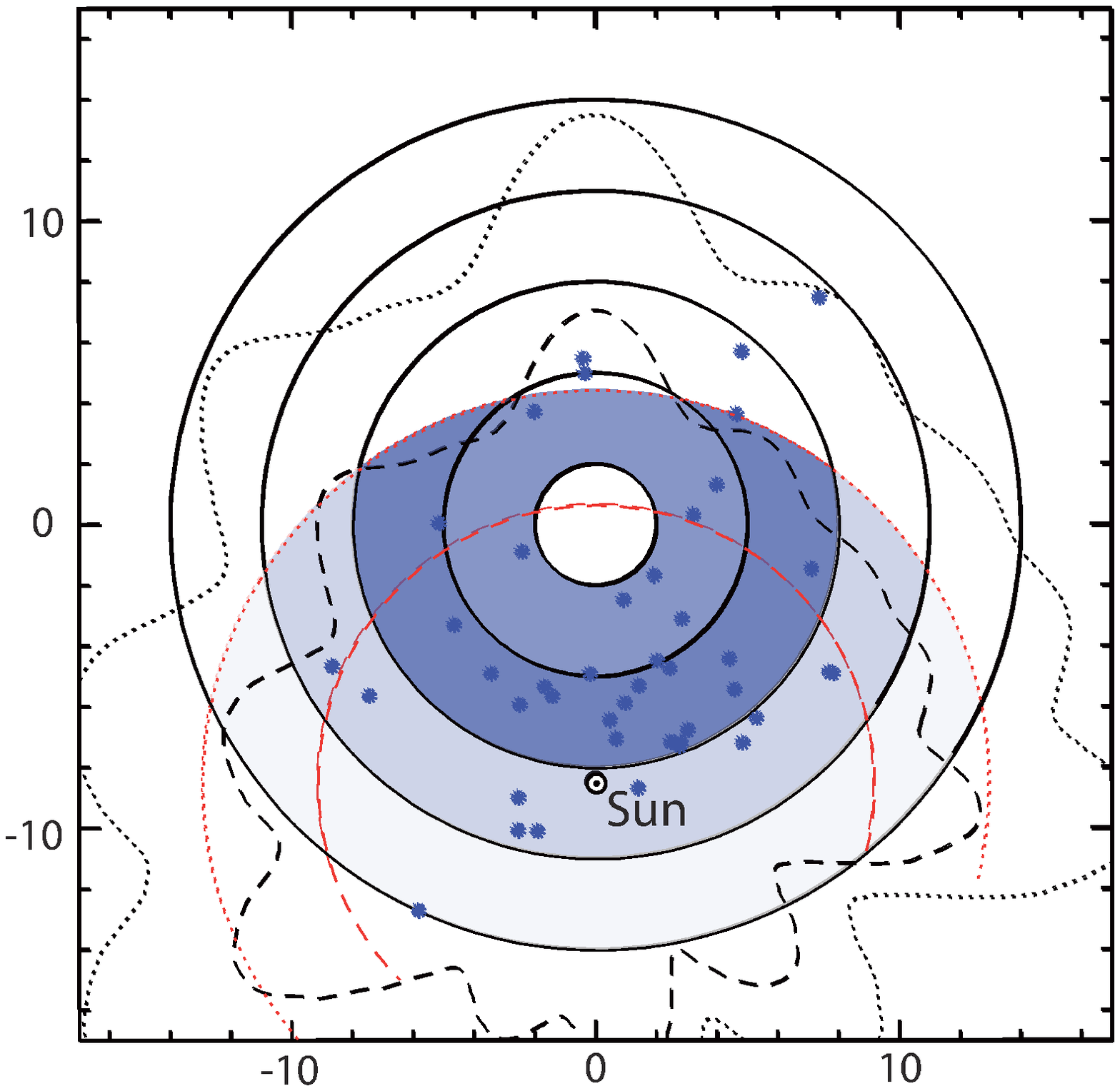}
\caption{An illustrative view of the surface density of HMXBs in the Galaxy (the darker color of the annulus corresponds to the higher surface density of HMXBs). Black dotted and dashed curves show areas of the Galaxy, within which the INTEGRAL Galactic survey detects all sources with luminosities $>10^{35.5}$~\ergs\ and $>10^{35}$~\ergs, respectively. Red dotted and dashed circles show distances, till which we detect all sources with luminosities higher than $10^{35}$~\ergs\ and  $2\times 10^{35}$~\ergs\ according to the adopted flux limit 0.7 mCrab in the inner part of the Galaxy $-107^\circ<l<136^\circ$. Blue points indicate positions of HMXBs from our sample. All distances are in kpc.}
\label{plotnosti}
\end{figure}

\begin{figure}
\includegraphics[width=\columnwidth,bb=28 180 575 695,clip]{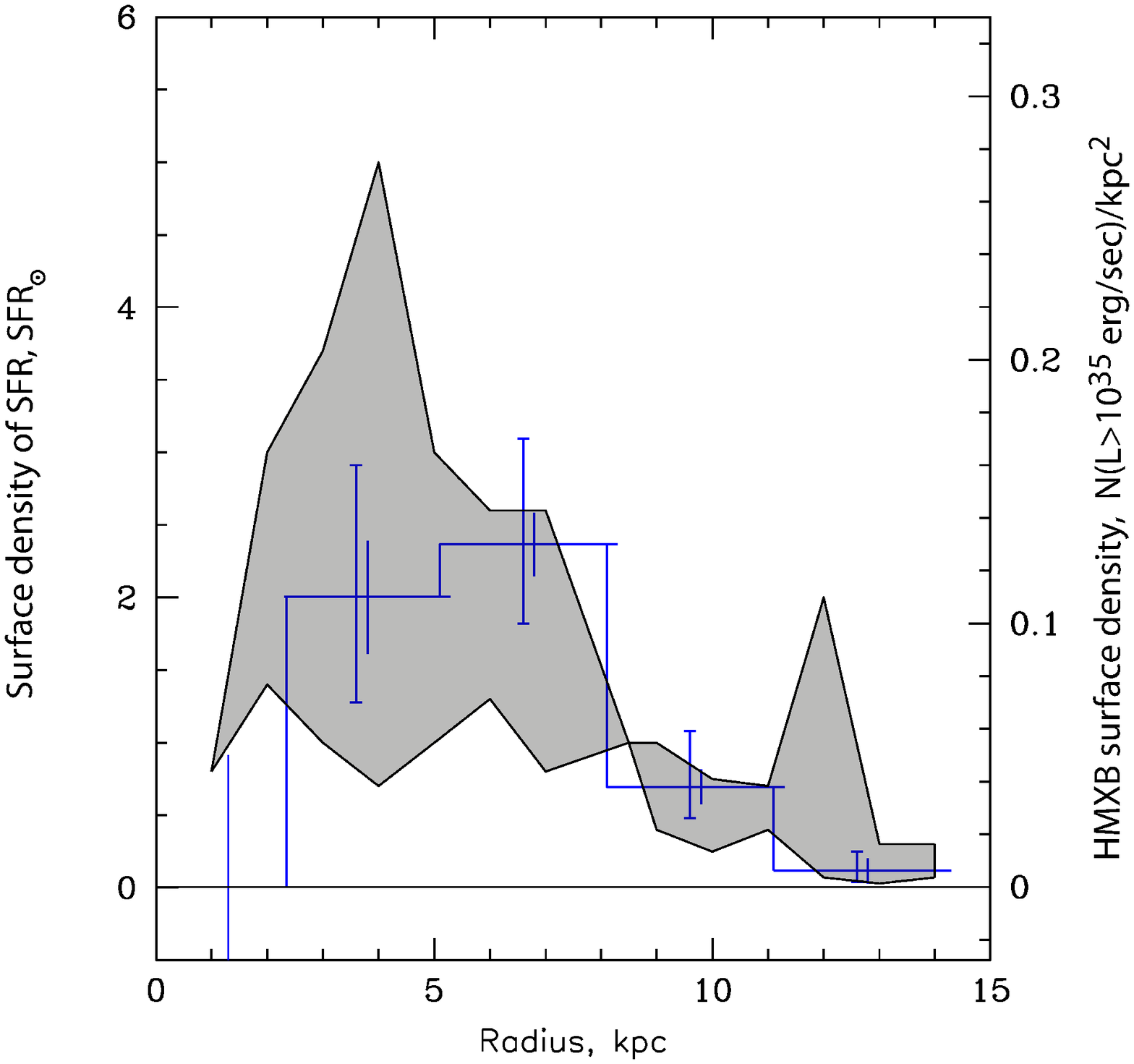}
\caption{Dependence of the HMXBs surface density (histogram, right axis) and star formation rate surface density (left axis) on the galactocentric distance. Star formation rates are presented by their upper and lower bounds (solid curves) from works \citet{guesten82,lyne85,chiappini01}.
Error bars on the histogram represent statistical (larger error bars) and systematic (smaller error bars) uncertainties. Systematic uncertainties are estimated by variations of distances to sources.}
\label{sfr_radius}
\end{figure}

\subsection{Vertical distribution}

We have fitted a vertical distribution of HMXBs in two commonly used ways: 1) with a simple exponential model of their volume density $\rho=\rho_0 \exp(-z/h)$ and 2) with a model of a self-gravitating isothermal disk $\rho=\rho_0 ~\textrm{sech}^2(z/\sqrt{2}h)$. The former model gives the best fit height $h=85^{+23}_{-15}$ pc, the latter one $h=90\pm 15$ pc. The systematic uncertainty of the scale-height due to a limited accuracy of distances to sources is smaller than the statistical one. This scale-height of the HMXBs distribution is somewhat smaller than that presented in papers of \citet{grimm02,dean05,2007A&A...467..585B}, likely due to the higher completeness and uniformity of our sample.

It is important to note that the scale-height of the HMXBs distribution is larger than the one of the distribution of massive stars in the Galaxy, e.g., the sample of WR stars has a scale of $\sim45$ pc \citep{conti90}, OB star formation regions $\sim 30$ pc \citep{bronfman00}, open clusters $\sim50$ pc \citep{pandey88,joshi05}. This indicates that HMXBs should have traveled a finite distance from their birth sites. If we will assume that HMXBs receive their systemic velocity during supernova explosions, we can make a rough estimate of the kinematic age of HMXBs after the supernova explosion \cite[see, e.g., similar estimates in][]{brandt95}.

Adopting the value $\sim$ 50-90 km\,$s^{-1}$ of the systemic velocity as a typical value for HMXBs \cite[see e.g.][]{kaper97,huthoff02} we can estimate their kinematic age, $\tau\simeq 50$ pc/$(50-90)$ km\,$s^{-1}$ $\simeq0.5-1$~Myr. We should emphasize here that these values are applicable to the wind-fed population, which we study in our sample. Roche lobe overflowing systems with luminosities above $(0.5-1)\times 10^{37}$ erg/sec might have smaller ages.

\section{Properties of the wind-fed NS HMXB population}

Having collected the statistically clear sample of wind-fed binaries with accreting neutron stars, we now can try to understand physical parameters which determine their population.

\subsection{Slice of HMXBs with fixed orbital periods}

The simplest picture of neutron stars accreting from a stellar wind was developed in classical works of \cite{hoyle40,bondi44,davidson73}. In this framework the luminosity of the accreting neutron star simply depends on the mass flow, intercepted by its gravitational field. Masses of neutron stars lie in the narrow range of values ($\sim1.3-1.9 M_\odot$), therefore they do not influence strongly on the HMXBs accretion luminosity. The mass $\dot{M}$, intercepted by a neutron star from the stellar wind of the companion mainly depends on three parameters: 1) the stellar wind mass loss rate $\dot{M}_{\rm w}$, 2) the distance between components in the binary system $a$ and 3) the stellar wind velocity $v_{\rm w}$ (here we adopt that the wind velocity is much larger than the velocity of the orbital motion of binary stars; such an assumption is valid for the fast wind of young massive stars of our sample). Roughly it can be expressed as

$$
\dot{M}\propto \dot{M}_{\rm w} v_{\rm w}^{-4} a^{-2}.~~~~~~~~~~~~~~~~~~~~~~~~~~~~~~~~~~ (3)
$$

The orbital separation in the binary system depends on its orbital period. For a fixed distance between components in HMXBs the X-ray luminosity of the neutron star will be a function of the mass loss rate and the wind velocity of the optical star. Speaking very generally, for some fixed distance between companions in the binary system its X-ray luminosity (the accretion powered luminosity of the neutron star) should depend mainly on the mass of the normal star. The larger mass of the optical star should lead to the higher X-ray luminosity of the binary. Lets compare these simple arguments with properties of the real sample.

\begin{figure}
\includegraphics[width=\columnwidth,bb=40 181 568 700,clip]{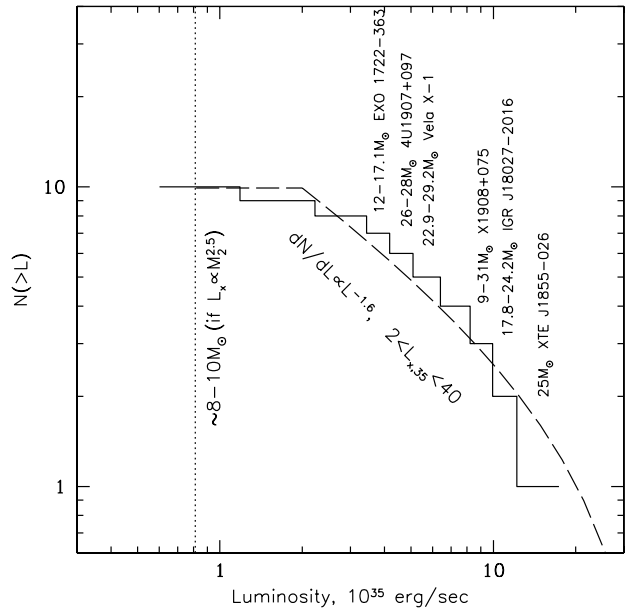}
\caption{Cumulative luminosity function of wind-fed NS-HMXBs from our sample with orbital periods in the range of $4-10$ days. Dashed curve demonstrates a model prediction calculated from a simple analytical formula $dN/dL\propto L^{-1.6}$ in the luminosity range $2.2\times10^{35}<L_x<8\times10^{36}$ \ergs. The companion star masses in some binaries are written along the histogram. They were taken from papers of \citet{2010ApJ...719..958T} for XTE\,J1855-026, \citet{2011A&A...532A.124M} for IGR\,J18027-2016, \citet{2004ApJ...617.1284L} for X\,1908+075, \citet{2003A&A...401..313Q} for \mbox{Vela\,X-1}, \citet{2005A&A...436..661C} for 4U\,1907+097, \citet{2010A&A...509A..79M} for EXO\,1722-363. Dotted line shows an approximate value of the X-ray luminosity of the HMXB with $M_2\sim8-10 M_\odot$ (see text for details).}
\label{slice4_10d}
\end{figure}

In Fig.\ref{slice4_10d} we present the cumulative luminosity function of the wind-fed HMXBs within a limited range of orbital periods from 4 to 10 days (the latter restricts somehow the distance between binary components).

From this figure it is clearly seen three main properties of this sub-sample:

\begin{itemize}
\item persistent HMXBs with orbital periods 4-10 days have luminosities in the range from $10^{35}$ \ergs\ to $2\times10^{36}$ \ergs\ and their luminosity function can be approximated by a power law $dN/dL\propto L^{-1.6}$ (dashed line);
\item there is a tentative trend that masses of optical stars (in cases where we can find them in the literature) decrease towards the lower X-ray luminosities;
\item the lowest luminosity of the HMXB in this small sample is $L_x\sim0.8\times10^{35}$ \ergs.
\end{itemize}

The power law shape of the X-ray luminosity function of HMXBs with fixed orbital periods was explained by \cite{postnov03} by properties of the mass distribution of secondaries in the binary, and scalings of their stellar wind mass loss rate.

An additional important hint from this plot is that for a given orbital period, it seems that there should be a minimal X-ray luminosity of the wind-fed HMXB.

This conclusion is not absolutely robust from the observational point of view because at luminosities below $L_x\sim(2-3)\times10^{35}$ \ergs\ our sample have a low completeness with the respect to orbital periods of binaries. Nevertheless, usually there are no significant observational problems to determine orbital periods in the range of 4-10 days, therefore we do not expect that this incompleteness is large.

From the theoretical point of view this lower boundary of the X-ray luminosity of the wind-fed HMXB can be easily understood. Let's consider the accretion onto the non-magnetic neutron star, i.e. without an additional 'accretion flow--neutron star magnetosphere' interaction, which can inhibit the accretion and thus significantly diminish the time average X-ray luminosity of such a binary system. If we will fix the orbital period of the binary system (that approximately corresponds to the fixed distance between companions), it will have the minimal X-ray luminosity if the optical star has the lowest mass. We can expect that the lowest masses in HMXBs should be approximately $8-10 M_\odot$. If we will use the 'X-ray luminosity -- mass of the optical companion' scaling from work of \citet{postnov03} $L_x \propto M_2^{2.5}$, we can estimate the minimal X-ray luminosity of the HMXB in this small sub-sample. The resulted minimal X-ray luminosity will be in a range of $(0.7-1)\times10^{35}$ \ergs, that well agrees with the distribution, presented in Fig.\ref{slice4_10d}.

\subsection{Toy model of the wind-fed NS-HMXB population}

\begin{figure}
\includegraphics[width=\columnwidth,bb=0 10 568 600,clip]{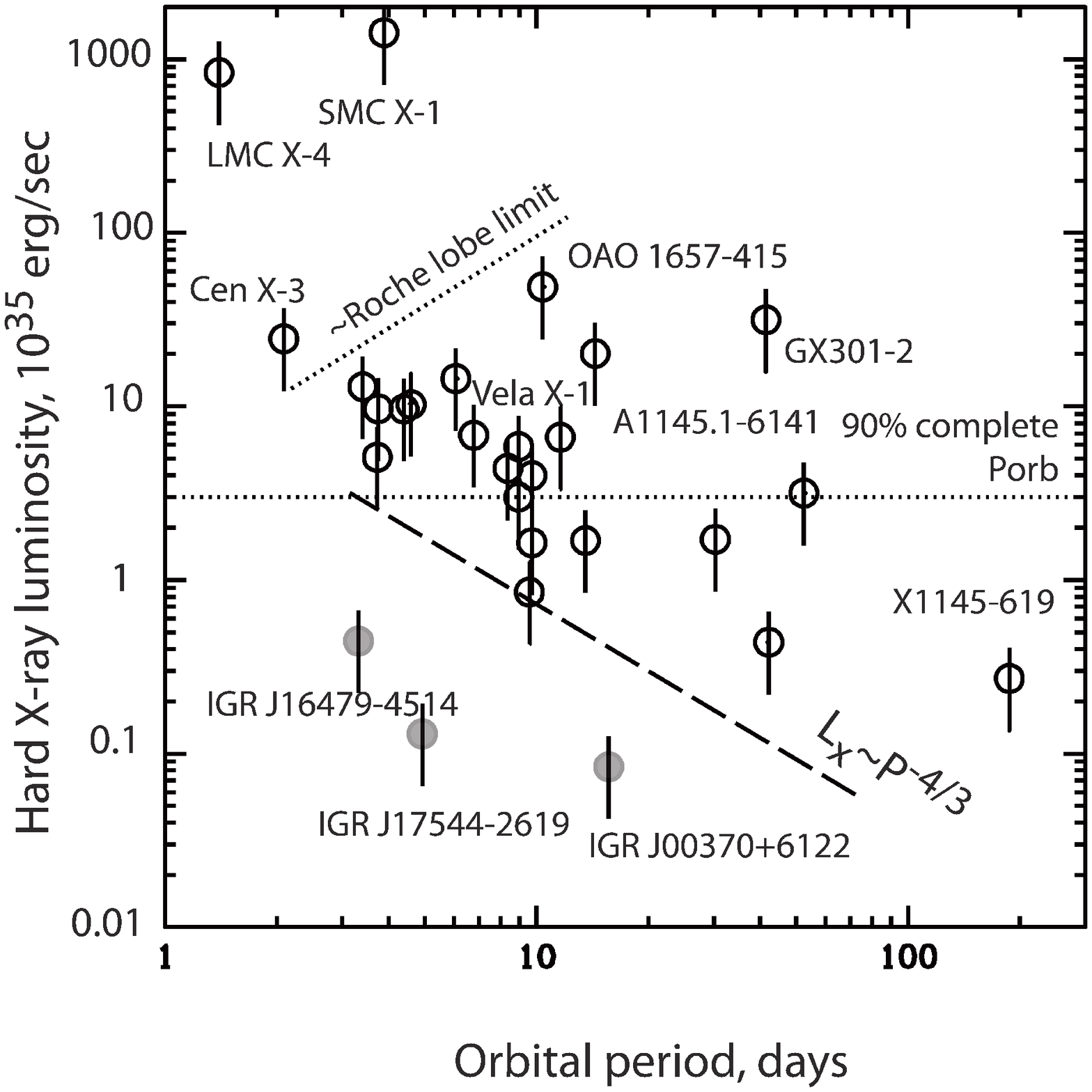}
\caption{Orbital periods and mean X-ray luminosities of HMXBs from the flux limited sample (open circles). Gray filled circles denote positions of known supergiant fast X-ray transients, which still passed our selection criteria for the ''persistency'' due to their faintness. In addition to sources in the Milky Way we also demonstrate positions of two HMXBs in Magellanic Clouds, which accrete via the Roche lobe overflow of the optical star. Dotted horizontal line shows the level above which we know orbital periods for more than 90\% of sources from our sample. Dashed line shows an approximate lower boundary of the ''allowed'' area for wind-fed accreting sources (see text).}
\label{scatter}
\end{figure}

In this section we will try to simulate the global properties of the population of NS-HMXBs, incorporating main ingredients from works of \cite{hoyle40,bondi44,davidson73,lamers76,iben95,postnov03,bhadkamkar12}.

The main simplification in the modeling of the HMXBs population comes from the fact that their lifetimes are small. Thus, it can be securely adopted that properties of these systems depend only on their current parameters and are not significantly influenced by the previous history. In our simple toy model we will use this fact explicitly.

As it was mentioned in Section 4.1 the X-ray luminosity of the neutron star fed by a fast stellar wind of a massive star is determined by: 1) the stellar wind density at the position of the compact object, and 2) the wind velocity.

The mass loss rate in the stellar wind $\dot{M}_{\rm w}$ is mainly a function of the mass of the optical star $M_2$ and the wind velocity $v_{\rm w}$ \cite[e.g.][]{castor75}.

$$
\dot{M}_{\rm w}\approx \epsilon {L_2\over{v_{\rm w}c}},~~~~~~~~~~~~~~~~~~~~~~~~~~~~~~~~~~~~~~ (4)
$$
\noindent
where $L_2$ is the luminosity of the optical star, which by-turn is connected with its mass $M_2$, $\epsilon$ a dimensionless efficiency parameter. It depends on the type and temperature of the optical star \citep{lamers76} and can be varied in the range of $\epsilon\simeq0.4-1.0$ for typical temperatures of supergiants \citep{chaty08,2008A&A...484..801R}. In following calculations we adopt $\epsilon\simeq0.6$.

For a fast stellar wind the mass accretion rate on the neutron star can be estimated as follows:

$$
\dot{M}\approx \dot{M}_{\rm w} \left({GM_{\rm ns}\over{v^2_{\rm w,ns}a}}\right)^2,~~~~~~~~~~~~~~~~~~~~~~~~~~~~~~~~~ (5)
$$
\noindent
where $v_{\rm w,ns}$ -- the wind velocity at the position of the neutron star $v_{\rm w,ns}\approx v_{\rm w}(1-R_2/a)^{1/2}$ \citep{castor75,vink00}, which might be significantly lower than the wind velocity at infinity $v_{\rm w}$ if the neutron star is not far from the optical star, $R_2$ is its radius.

For the mass-radius relation for optical stars in the binary system we adopt ${R_2/R_\odot}\simeq0.9 M_2/{M_\odot}$, which was calculated for sources in our sample. It approximately true in general (but with a large scatter due to evolutionary effects). For the mass-luminosity relation we adopt $L_2/L_\odot=19(M_2/M_\odot)^{2.76}$ \citep{vitrichenko07}.

Combining above formulas with the assumed mass to energy conversion during the neutron star accretion $L_x=0.1\dot{M}c^2$,
we can simply estimate that the X-ray luminosity of the wind-fed neutron star is:

\begin{eqnarray}
L_x\approx 5.4 k \times 10^{35} \left({M_2\over{10~M_\odot}}\right)^{2.76}\times \nonumber \\
 \times \left({a\over{10~R_\odot}}\right)^{-2}
\left( {v_{\rm w}[1-R_2/a]^{1/2}\over{1000~\textrm{km $s^{-1}$}}}\right)^{-5}   \nonumber
\end{eqnarray}
\hfill(6)

\medskip
\noindent
where coefficient $k$ incorporates all uncertainties in the approach used for the hard X-ray luminosity estimation. To describe the observed population of HMXBs (see Fig.\ref{scatter},\ref{lf_sim}) in the best way this coefficient should be around $\simeq1.4-1.5$. Taking into account a number of suggestions and simplifications in our model it can be considered as a reasonable value.

The line of reasoning, enlisted above, predicts that persistently accreting wind-fed neutron stars should occupy some particular region on a scatter plot $P_{\rm orb}-L_x$ (see Fig.\ref{scatter}). Indeed, at any given orbital period $P_{\rm orb}$ (which can be translated into the orbital separation) there should be a lower limit on the HMXB X-ray luminosity, corresponding (within the framework of our simple consideration) to the binary with the smallest masses of the companion star and thus to the lowest values of the mass loss rate. This lower bound should have a functional form $L_{\rm x}\propto a^{-2}\propto P_{\rm orb}^{-4/3}$ at large orbital periods and, roughly speaking, divide the $P_{\rm orb}-L_x$ diagram into two areas -- ''allowed'' for wind-fed neutron stars and ''forbidden'' for them.
It is remarkable that we do see this lower boundary on the diagram $P_{\rm orb}-L_x$, shown on Fig.\ref{scatter} (dashed line). This provides a significant support for our simple approach.

\subsection{Luminosity function of the simulated population of wind-fed HMXBs}

\begin{figure}
\includegraphics[width=\columnwidth,bb=36 181 568 700,clip]{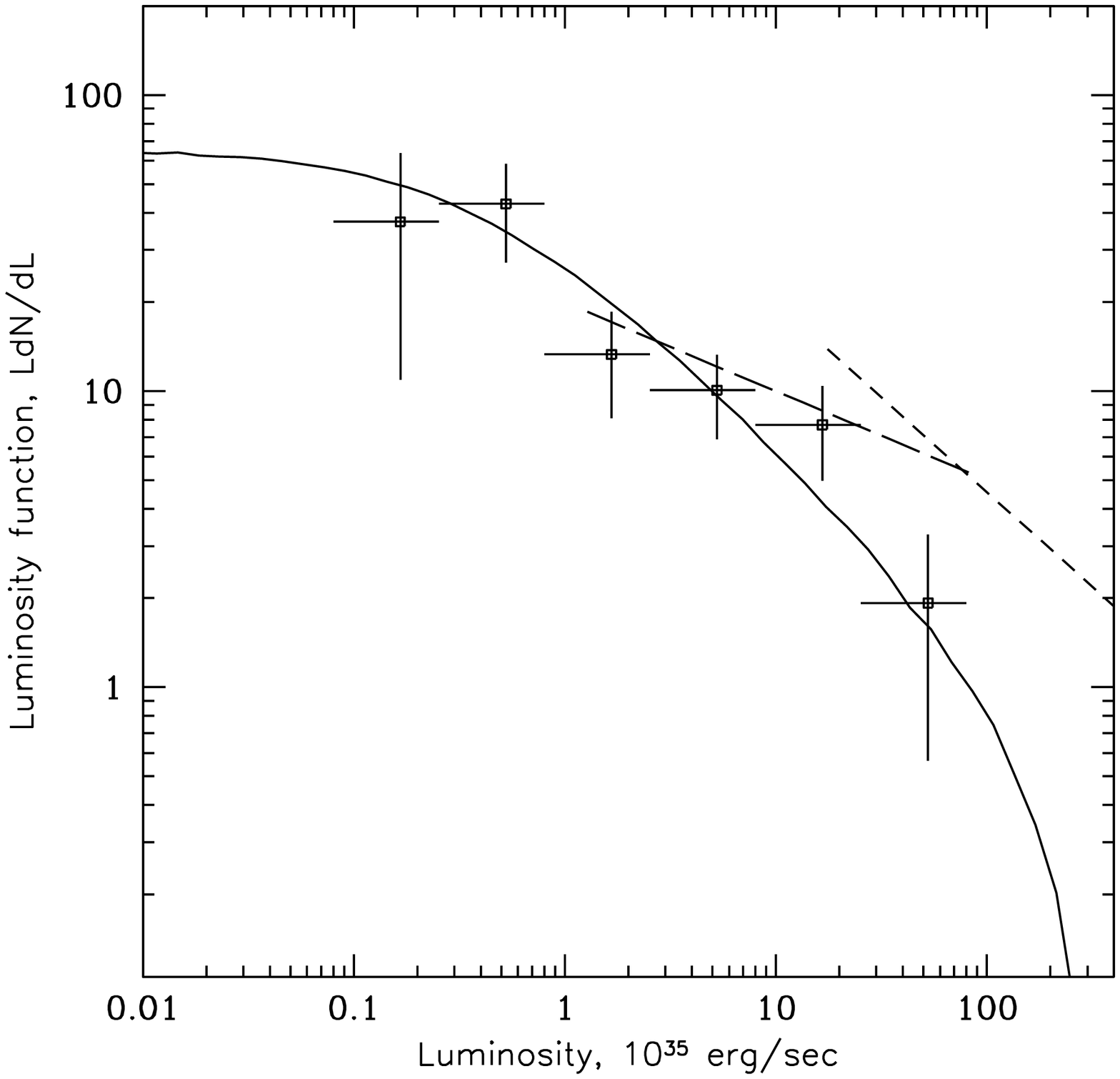}
\caption{Luminosity function of persistent HMXBs in the Milky Way as seen by INTEGRAL (crosses) along with the model of the simulated HMXBs population (solid line). Long-dashed and short-dashed lines show best fit models of the HMXBs LF obtained by \citet{voss10} and \citet{grimm02}, respectively. The former one was calculated in the $15-55$ keV energy band, where the luminosity of NS-HMXBs is practically the same as the one on the $17-60$ keV energy band. Therefore the corresponding LF is presented in its respective luminosity interval. The LF in the paper of \citet{grimm02} was calculated in the $2-10$ keV energy band. Therefore, to recalculate it to the $17-60$ keV energy band we used an approximate ratio between fluxes $F_{\rm 2-10 keV}/F_{\rm 17-60 keV}\simeq0.5$, which was derived from the analysis of broadband spectra of typical HMXBs with neutron stars (see, e.g., \citealt{filippova05}).}
\label{lf_sim}
\end{figure}

In order to simulate the population of HMXBs with neutron stars we need to assume some distribution of masses of secondaries and orbital periods of these systems. \cite{postnov03} suggested that secondary stars in HMXBs have an initial (Salpeter) mass distribution ignoring a possible evolution of this distribution before the beginning of an active HMXB phase. \cite{bhadkamkar12} recently analyzed this evolution and showed that really the overall power law slope of this distribution does not change significantly in the mass interval $15-40M_\odot$. Therefore, for our subsequent modelling we adopt the mass distribution of secondaries in the form ${dN/dM}\propto M^{-2.35}$ in the mass interval 10-60 $M_\odot$.

For logarithm of orbital periods we assume a flat distribution \cite[see e.g.][]{opik24,masevich88}, but restrict this distribution by the interval $1.3<\log P(\textrm{days})<3.3$ with gaussian tails of the width 0.4 in order to mimic results of the pre-HMXB evolution of binaries\citep{bhadkamkar12}.

It is necessary to emphasize here that in our toy model we not take into account several effects, which might be important for the detailed comparison of properties of the simulated HMXBs population with the observed one. In particular:

\begin{itemize}
\item we adopt some specific mass-radius relation for optical stars; this assumption led to the fact that the wind velocity at the infinity $v_{\rm w}$ is the same for all stars in the simulated population; an evolution of young massive stars can lead to a spread (at least) in this value;
\item we do not take into account that an X-ray illumination of the wind matter from the accreting neutron star can lead to a deviation of the wind flow from a simple analytic formula (4);
\item we do not take into account that some part of the persistent NS-HMXBs population can accrete a matter from the dense equatorial wind of Be stars (i.e. X\,Per or X\,1145-619); it will raise their X-ray luminosities above what might be expected for a binary system with a spherically symmetric stellar wind;
\item we do not take into account that depending on the magnetic field and spin period of the accreting neutron star the accretion flow might be stopped by a propeller mechanism \citep{illarionov75}; it should lead to a disappearance of binaries from different parts of our $P_{\rm orb}-L_{\rm x}$ diagram and can distort the shape of the luminosity function.
\end{itemize}

Nevertheless, below we show that in spite of these limitations our toy model reasonably well describe observational appearances of HMXBs.

A comparison of the luminosity function of the simulated wind-fed HMXBs population with that measured for such systems in our Galaxy is presented in Fig.\ref{lf_sim}. It is clearly seen that in spite of our simplistic approach to the simulation of the HMXBs population their luminosity function closely follows the observed LF of wind-fed NS-HMXBs in the Milky Way. At the same time, the measured luminosity function of HMXBs is somehow different from those of \citet{grimm02} and \citet{voss10} (dotted and dashed lines, respectively). The origin of this difference can be connected with the absence of black hole accretors, Roche-lobe filling systems and transients in our sample, and another way of the correction for the incompleteness.

The main properties of the luminosity function of the HMXBs population are the steep cutoff at luminosities above $\sim 10^{36}$ \ergs\ and its flattening at luminosities below $\sim 10^{34-34.5}$ \ergs. According to our simple population model it is likely that wind-fed HMXBs are limited by a hard X-ray luminosity $\sim5\times10^{36}$ \ergs. This conclusion is not new and was mentioned already by e.g. \citet{lamers76}. More luminous HMXBs should have a different nature. For example, they can harbor WR stars with more powerful winds, can accrete from Roche lobe overflowing massive giants (which can provide a mass accretion rate as high as $10^{-4} M_\odot$ yr$^{-1}$ or more), can have black holes as primaries, etc. In our Galaxy we know only a few such sources (e.g., Cyg X-1, Cyg X-3), which, being taken together with wind-fed NS-HMXBs, will make the bright part of the HMXBs LF flatter than we see it on Fig.\ref{lf_sim}. This fact could be a reason why the luminosity function of HMXBs in galaxies with high star formation rates continues with a power law $d\log N/d\log L_{\rm x}\sim-1.6$ towards luminosities much higher than in our sample \citep{grimm03,mineo12}. The flat part of the LF at low luminosities (below $\sim 10^{34}$ \ergs) is a consequence of a log-constant distribution of orbital periods of binary systems. But, it is necessary to note, that at the present level of the survey sensitivity this flattening is not very significant.

\begin{figure}
\includegraphics[width=\columnwidth,bb=130 105 415 385,clip]{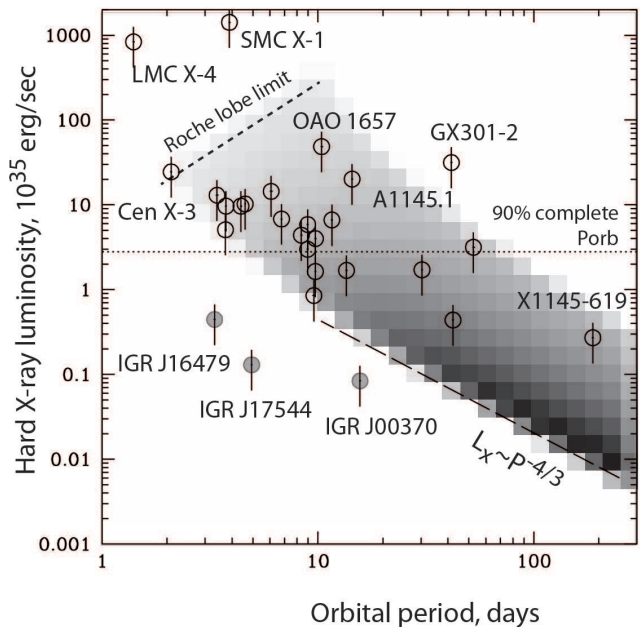}
\caption{Density of sources, produced with our toy model of the wind-fed NS-HMXBs population (gray scale) along with the distribution of known galactic sources on the plane $P_{\rm orb}-L_{\rm x}$ (same as Fig.\ref{scatter}).}
\label{scatterplot_grayscale}
\end{figure}

The corresponding distribution of the simulated HMXBs population on the $P_{\rm orb}-L_{\rm x}$ diagram is shown in Fig.\ref{scatterplot_grayscale} by a gray scale. It is clearly seen that it forms the ''allowed'' region on the diagram, where most of persistent HMXBs are located.

\subsection{Supergiant fast X-ray transients}

Monitoring observations of the Galaxy with the INTEGRAL observatory have revealed a new pattern of a variability of more than dozen of high mass X-ray binaries -- short bright flares with long periods of quiescence \citep{smith98,sunyaev03,sguera05,negueruela06,sidoli11}. These sources rapidly have grown up to be a special subpopulation of HMXBs  -- supergiant fast X-ray transients (SFXTs, virtually all of them shown to be containing early type supergiant companions).

An origin of such flares is not yet fully understood. Proposed models can be separated roughly in two main branches: a) flares occurring due to an occasional accretion of blobs of the matter from the clumpy stellar wind of a supergiant \cite[see, e.g.,][]{walter07,oskinova12}, and b) magnetic arrest of the accretion due to a rotating neutron star magnetosphere and its occasional breakthrough \cite[see, e.g.,][]{grebenev07,bozzo08,postnov08,shakura12}. Some combination of these two types of models was considered by \citet{ducci10}.

In order to reveal a true nature of the SFXTs phenomenon it is very informative to put these sources in a context of the outlined toy model of the wind-fed HMXBs population.

\begin{figure}
\includegraphics[width=\columnwidth,bb=20 140 555 649,clip]{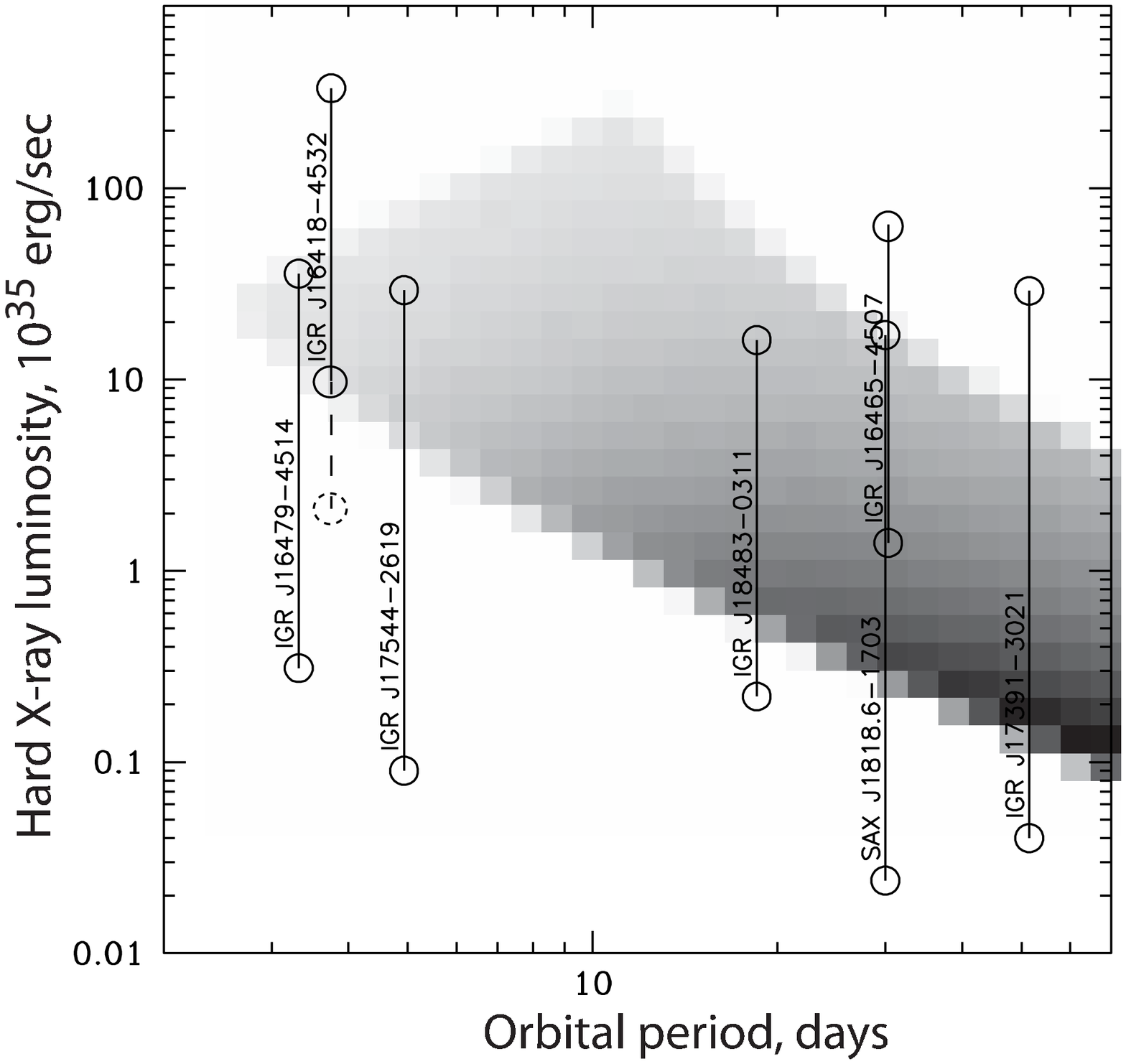}
\caption{Positions of supergiant fast X-ray transients (SFXTs) on the $P_{\rm orb}-L_{\rm x}$ diagram. Here all sources are shown by two values -- using the median and maximal fluxes, measured by the INTEGRAL observatory (solid circles). Dashed circle denotes the measurements of the minimal flux for IGR\,J16418-4532 based on the XMM-Newton data \citep{sidoli12} and recalculated to the 17-60 keV energy band.}
\label{porb_lx_sfxt}
\end{figure}
\begin{table*}

   \caption{Luminosities of some Supergiant Fast X-ray Transients.$^a$\label{tab:sfxtpar}}
   \hspace{-7mm}\begin{tabular}{l|r|r|r|r|r}
     \hline
     \hline
  Name                    & $L_{X, med}$,$^b$     & $L_{X, peak}$,$^c$      & Distance, & $P_{orb}$,  & References$^{d}$ \\
                          & $10^{35}$ erg s$^{-1}$ & $10^{35}$ erg s$^{-1}$ & kpc       & days &            \\
      \hline

  IGR\,J16418-4532   & 9.8  &  334.2&   13.0  & 3.75   &  1, 2  \\
  IGR\,J16465-4507   & 1.4  &  74.6 &    9.4  & 30.32  &  3, 4  \\
  IGR\,J16479-4514   & 0.31 &  35.8 &    2.8  & 3.32   &  3, 5  \\
  IGR\,J17391-3021   & 0.04 &  29.1 &    2.7  & 51.47  &  3, 6 \\
  IGR\,J17544-2619   & 0.09 &  29.4 &    3.6  & 4.93   &  7, 8 \\
  SAX\,J1818.6-1703  & 0.024&  17.1 &    2.1  & 30.0   &  9, 10 \\
  IGR\,J18483-0311   & 0.22 &  16.1 &    2.8  & 18.55  &  9, 11 \\

    \hline
    \end{tabular}
    \medskip
    \hspace{-17mm}\begin{tabular}{ll}
    $^a$ & in the 17--60 keV energy band \\
    $^b$ & most probable (median) luminosity \\
    $^c$ & maximum luminosity, averaged over $\sim2$ ksec observations \\
    $^d$ & references correspond to the distance and orbital period measurements\\
    \end{tabular}\\

    \medskip

\parbox{17.5cm}
{\tiny
References:
(1) \cite{2008A&A...484..783C},
(2) \cite{2006ATel..779....1C},
(3) \cite{2008A&A...484..801R},
(4) \cite{2010MNRAS.406L..75C},
(5) \cite{2009MNRAS.397L..11J},
(6) \cite{2010MNRAS.409.1220D},
(7) \cite{2006A&A...455..653P},
(8) \cite{2009MNRAS.399L.113C},
(9) \cite{2010A&A...510A..61T},
(10) \cite{2009A&A...493L...1Z},
(11) \cite{2007A&A...467..249S}.
}
\end{table*}

A sub-sample of SFXTs (and SFXT candidates) with known distances and orbital periods is presented in Table \ref{tab:sfxtpar} and shown in Fig.\ref{porb_lx_sfxt}. Now every source has two values of its luminosity -- the most probable (median) value, calculated using all INTEGRAL measurements, and the peak luminosity value. The last ones were extracted from corresponding light curves of sources binned into $\sim2$ ksec bins. It should be noted here that the true peak luminosities might be higher in some cases if smaller time bins would be used.

It is clear that the median luminosities of genuine SFXTs lie in the forbidden area, i.e. below the line of the minimal X-ray luminosity, allowed for the wind-fed NS in the binary system with the young star having the smallest mass loss rate in the wind (see Section 4.2).
{\sl It explicitly means that most of the time the accretion in SFXTs is inhibited by some mechanism.}
The only one source, which median value lies in the allowed area -- IGR\,J16465-4507 -- sometimes classified as an "intermediate" SFXT, and its flux variability is largely related with the orbital modulation of the accretion flow \citep{2010MNRAS.406L..75C}. Somewhat a similar situation is emerging for IGR\,J16418-4532, which is firmly established SFXT, but located near the border between allowed and forbidden areas. It happen only because we were limiting ourself with data of INTEGRAL/IBIS, which are significantly worse that those of focusing X-ray telescopes. If we instead use the results of \citet{sidoli12}, obtained with the XRT/Swift telescope, the median luminosity of the source in the quiescent state will be $\sim2\times10^{35}$ \ergs, that is $\sim5$ times lower than it was measured by INTEGRAL. In this case the source position in the quiescence state will be deeply in the forbidden area (dashed line and circle in Fig.\ref{porb_lx_sfxt}). Finally, concerning IGR\,J16418-4532, it is necessary to note a possible high uncertainties for its distance.

The clumpy wind model can produce flares, but it can not explain why the median X-ray luminosity of the SFXT binary is much lower than that of the similar binary with the similar orbital period, but not classified as the SFXT.

In order to see directly the outlined dichotomy we can compare sources IGR\,J17544-2619 and IGR\,J18027-2016. Optical stars in both binaries have similar spectral types -- O9Ib for the former and OBsg for the latter, both have the similar orbital periods (around 5 days) and thus the similar separation between components in the system. But the first one is a fast transient (SFXT) and the latter one is a normal accreting pulsar. This example clearly shows that there are some additional parameters in the binary system (e.g., magnetic field and spin period of the neutron star), which determines its global behavior.

The position of the SFXTs binary along the Y-axis in Fig.\ref{porb_lx_sfxt} should depend on the duty cycle of its activity. It is natural to propose that the duty cycle should be higher and the ratio of the peak fluxes to the median ones should be smaller if the source is closer to the allowed region. Such a source will "degenerate" into the normal persistent supergiant HMXB if its position will be within the allowed region. Note, that systematic studies of known SFXTs do show such a trend for all sources except for IGR\,J16479-4514 \citep{romano09,sidoli11}. It is possible that such a deviation in the overall trend is related with the fact that in this binary the optical star is close to filling its Roche lobe and our toy model of the wind accretion is not applicable to it.

\section{Faintest sources of the INTEGRAL survey and beyond}

A general understanding of the luminosity function of HMXBs and their Galaxy-wide distribution provides us a possibility to make predictions of a number of persistent sources at fainter fluxes.

Using the real sensitivity of the INTEGRAL Galactic plane survey instead of our adopted value 0.7 mCrab (see Fig.\ref{sensflux}) we can predict that in total we should detect up to $\sim 54$ persistent HMXBs (here we have taken into account that surface densities, calculated in Section 3.2, do not include 5 HMXBs with unknown distances and thus the true surface densities of HMXBs might be $\sim$14\% higher than it is presented in Table 2). Forty two of them were already observed with the adopted flux limit 0.7 mCrab. Below this limit there should be only $\sim$12 persistent HMXBs among all unidentified INTEGRAL sources. Six of them were already detected in the survey (see Table \ref{tab:srclist}). Judging from estimates of surface densities of different types of sources on the sky we can anticipate that the majority of the remaining galactic sources will be cataclysmic variables (CVs).

\begin{figure}
\includegraphics[width=\columnwidth,bb=20 239 571 755,clip]{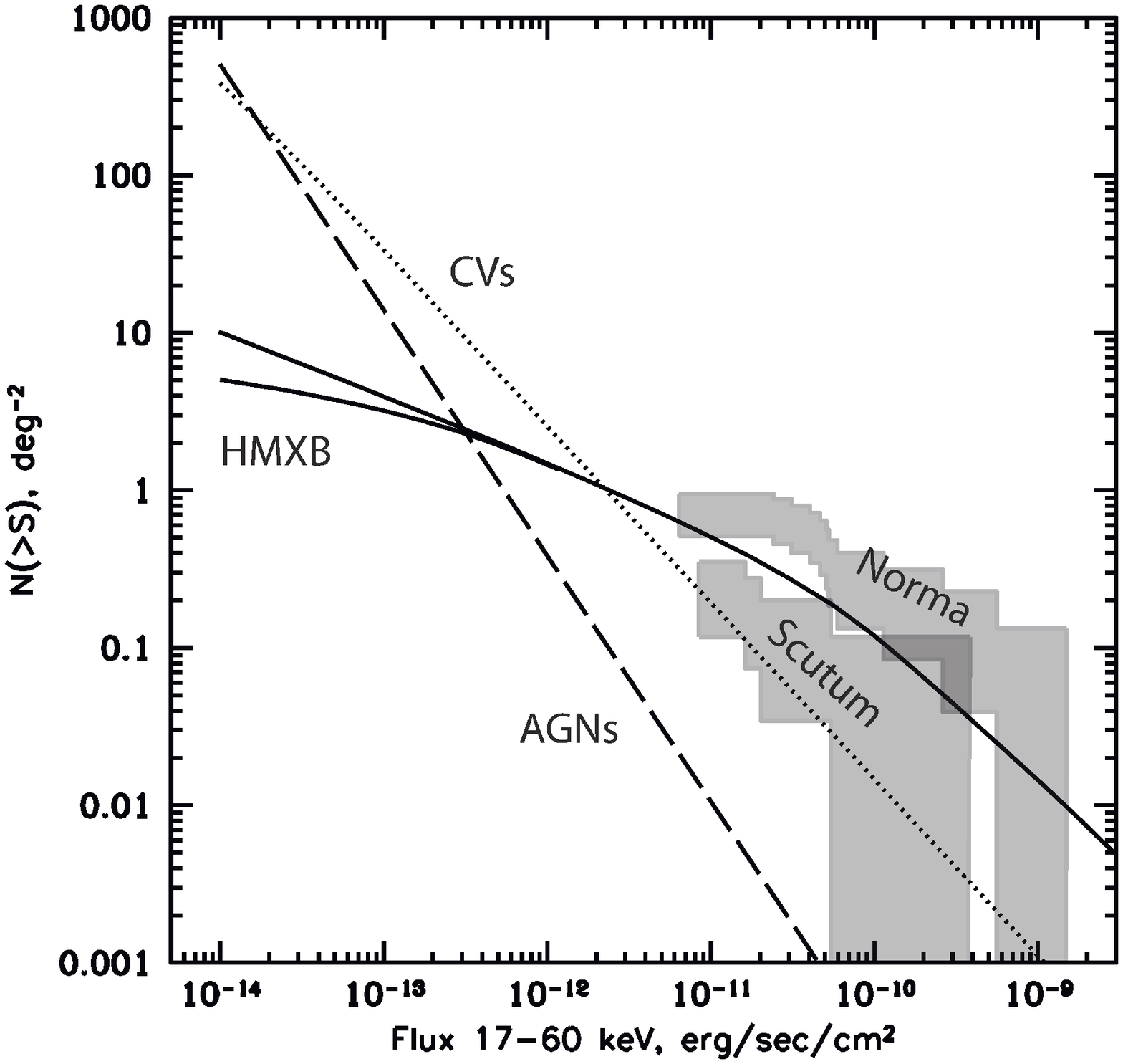}
\caption{Surface density of HMXBs in the direction to the Norma ($345^\circ<l<330^\circ$) and Scutum regions ($15^\circ<l<35^\circ$) in the Galactic plane (gray areas). Lines show predictions of numbers of different types of sources in these areas: AGNs (dashed line), CVs (dotted line) and HMXBs (solid lines). Predictions of a number of HMXBs at faint fluxes were done using two types of their LF with different slopes $d\log N/d\log$ at luminosities below $10^{34}$ \ergs\ (at higher luminosities the LF was taken as measured, see Table 2): 1) an upper solid curve -- the slope is fixed at the value $-1.4$, as measured at $10^{34}$ \ergs$<L_{\rm x}<10^{36}$ \ergs\, 2) a lower solid curve -- the slope of the LF is fixed at the value $-1$, as predicted from our toy model (see Fig.\ref{lf_sim}).}
\label{lognlogs_predict}
\end{figure}

This conclusion follows from Fig.\ref{lognlogs_predict}, where the expected number of HMXBs in flux limited surveys of the Galacti plane is shown in line with ones for active galactic nuclei (AGNs) and CVs. A number-flux function of AGNs was taken from \cite{krivonos10}, a number-flux function of cataclysmic variables was calculated using the parametrization from \cite{revnivtsev08b}. It is clear that the number of CVs and AGNs begin to dominate over HMXBs at fluxes below $\sim 10^{-12}$ \flux, which will be achieved with the new generation of hard X-ray telescopes like NuSTAR \citep{harrison10} and Astro-H \citep{takahashi10}. Small fields of view of these instruments will limit the detection rate of HMXBs. In order to increase the number of known persistent HMXBs in the Galaxy (mainly due to low luminosity sources) one need to use large survey missions like Spectrum-RG \citep{pavlinsky09}.

\section{Summary}
We have constructed for the first time a well defined sample of non-transient High Mass X-ray Binaries in our Galaxy using the flux limited 9-years long survey of the INTEGRAL observatory. Our results can be summarized as follows:

\begin{itemize}
\item The majority of persistent HMXBs accrete matter from the stellar wind of their supergiant/giant companions;
\item The luminosity function of wind-fed persistent HMXBs with accreting neutron stars can be described by a broken power law with the break around  $\sim2\times 10^{36}$ \ergs; at high luminosities the power law slope of the differential luminosity function is $\gamma_{\rm bright}>2.2$, at low luminosities $\gamma_{\rm faint}\approx 1.4$;
\item Using this luminosity function we have showed that the predicted number of HMXBs will be significantly lower than number of CVs and AGNs in future surveys with the better sensitivity (NuSTAR, Astro-H);
\item The spatial density distribution of wind-fed NS-HMXBs over the Galaxy have been measured. We have divided the Galaxy into several annuli and showed that the HMXBs surface density has a maximum on galactocentric distances 2-8 kpc. Such a distribution correlates well with the distribution of the surface density of the star formation rate in the Galaxy.
\item We have constructed a simple toy model of the wind-fed NS-HMXBs population and showed that it can adequately describe properties of the observed population. The model clearly shows that wind-fed HMXBs should disappear at luminosities higher than $\sim5\times10^{36}$ \ergs\ and their LF should flatten at $L_{\rm x}<10^{34}$ \ergs. The overall shape of the luminosity function of the simulated population agrees with the measured one;
\item We argued that for wind-fed NS-HMXBs there is an "allowed" region on the $P_{\rm orb}-L_{\rm x}$ diagram due to general properties of the wind accretion. We demonstrated that all persistent wind-fed HMXB lies in this allowed area.
\item All supergiant fast X-ray transients (SFXTs) in their quiescent state lie in the ''forbidden'' area of the $P_{\rm orb}-L_{\rm x}$ diagram and only during flares they ''jump'' into the ''allowed'' region. This strongly supports the idea that their transient behavior is caused by some kind of (possible magnetic) inhibition of the accretion onto the neutron star.
\end{itemize}

\section*{Acknowledgements}
Authors are grateful to Jerome Rodriguez and Arash Bodaghee for their efforts to support the web-site dedicated to IGR sources (http://irfu.cea.fr/Sap/IGR-Sources/). AL and MR thank Marat Gilfanov, Sergey Sazonov and Arash Bodaghee for valuable discussions of methods and results. Authors also thanks to the anonymous referee for the careful reading of the manuscript and useful comments. We would like to thank E.M.\,Churazov for developing the IBIS data analysis algorithms and providing the software. This work was supported by the grants of President of Russian Federation (RF) MD-1832.2011.2, NSh-5603.2012.2, grants RFBR 10-02-00492, 12-02-01265, programs P21 and OFN17 of Russian Academy of Sciences (RAS), State contract 14.740.11.0611 and grants 8701 and 8629 from Ministry of Science and Education of RF.

\label{lastpage}

\end{document}